\documentclass[twocolumn]{aastex631}

\usepackage{multirow}
\usepackage{wasysym}
\usepackage{enumerate}
\usepackage{amssymb} 
\shorttitle{A Giant Planet Candidate Orbiting WD\,0310}
\shortauthors{Limbach et al.}
\graphicspath{{./}{figures/}}

\begin{document}
\title{The MIRI Exoplanets Orbiting White Dwarfs (MEOW) Survey: Mid-Infrared Excess Reveals a Giant Planet Candidate around a Nearby White Dwarf
}

\correspondingauthor{Mary Anne Limbach}
\email{mlimbach@umich.edu}

\author[0000-0002-9521-9798]{Mary Anne Limbach}
\affiliation{Department of Astronomy, University of Michigan, Ann Arbor, MI 48109, USA}

\author[0000-0001-7246-5438]{Andrew Vanderburg}
\affiliation{Department of Physics and Kavli Institute for Astrophysics and Space Research, Massachusetts Institute of Technology, Cambridge, MA 02139, USA}

\author[0000-0002-8400-1646]{Alexander Venner}
\affiliation{Centre for Astrophysics, University of Southern Queensland, Toowoomba, QLD 4350, Australia}

\author[0000-0002-9632-1436]{Simon Blouin}
\affiliation{Department of Physics and Astronomy, University of Victoria, Victoria, BC V8W 2Y2, Canada}

\author[0000-0002-7352-7941]{Kevin B. Stevenson}
\affiliation{Johns Hopkins APL, 11100 Johns Hopkins Rd, Laurel, MD 20723, USA}

\author[0000-0003-4816-3469]{Ryan J. MacDonald}
\altaffiliation{NHFP Sagan Fellow}
\affiliation{Department of Astronomy, University of Michigan, Ann Arbor, MI 48109, USA}

\author[0000-0001-9827-1463]{Sydney Jenkins}
\affiliation{Department of Physics and Kavli Institute for Astrophysics and Space Research, Massachusetts Institute of Technology, Cambridge, MA 02139, USA}

\author[0000-0001-5831-9530]{Rachel Bowens-Rubin}
\affiliation{Department of Astronomy, University of Michigan, Ann Arbor, MI 48109, USA}

\author[0000-0001-7493-7419]{Melinda Soares-Furtado}
\altaffiliation{NASA Hubble Science Fellow}
\affiliation{Department of Astronomy,  University of Wisconsin-Madison, 475 N.~Charter St., Madison, WI 53706, USA}

\author[0000-0002-4404-0456]{Caroline Morley }
\affiliation{Department of Astronomy, University of Texas at Austin, Austin, TX, USA}

\author[0000-0001-8345-593X]{Markus Janson}
\affiliation{Department of Astronomy, Stockholm University, AlbaNova University Center, 10691 Stockholm, Sweden}

\author[0000-0002-1783-8817]{John Debes}
\affiliation{AURA for the European Space Agency (ESA), ESA Office, Space Telescope Science Institute, 3700 San Martin Drive, Baltimore, Maryland 21218, USA}

\author[0000-0002-8808-4282]{Siyi Xu}
\affil{Gemini Observatory/NSF's NOIRLab, 670 N. A'ohoku Place, Hilo, Hawaii, 96720, USA}

\author[0009-0004-8080-5358]{Evangelia Kleisioti}
\affiliation{Leiden Observatory, Leiden University, PO Box 9513, 2300 RA Leiden, The Netherlands}
\affiliation{Aerospace Engineering, TU Delft, Building 62 Kluyverweg 1, 2629 HS Delft, The Netherlands}

\author[0000-0002-7064-8270]{Matthew Kenworthy}
\affiliation{Leiden Observatory, Leiden University, PO Box 9513, 2300 RA Leiden, The Netherlands}

\author[0000-0003-1305-3761]{Paul Butler}
\affiliation{Earth and Planets Laboratory, Carnegie Institution for Science, 5241 Broad Branch Road, NW, Washington, DC 20015, USA}

\author[0000-0002-5226-787X]{Jeffrey D. Crane}
\affiliation{The Observatories of the Carnegie Institution for Science, 813 Santa Barbara Street, Pasadena, CA 91101, USA}

\author[0000-0003-0412-9664]{Dave Osip}
\affiliation{Las Campanas Observatory, Carnegie Institution for Science	Colina el Pino, Casilla 601 La Serena, Chile}

\author[0000-0002-8681-6136]{Stephen Shectman}
\affiliation{The Observatories of the Carnegie Institution for Science, 813 Santa Barbara Street, Pasadena, CA 91101, USA}

\author[0009-0008-2801-5040]{Johanna Teske}
\affiliation{Earth and Planets Laboratory, Carnegie Institution for Science, 5241 Broad Branch Road, NW, Washington, DC 20015, USA}

\begin{abstract}
The MIRI Exoplanets Orbiting White dwarfs (MEOW) Survey is a cycle 2 JWST program to search for exoplanets around dozens of nearby white dwarfs via infrared excess and direct imaging. In this paper, we present the detection of mid-infrared excess at 18 and 21 microns towards the bright ($V$~=~11.4) metal-polluted white dwarf WD 0310-688. The source of the IR excess is almost certainly within the system; the probability of background contamination is $<0.1\%$. While the IR excess could be due to an unprecedentedly small and cold debris disk, it is best explained by a $3.0^{+5.5}_{-1.9}$\,M$_{\rm Jup}$ cold (248$^{+84}_{-61}$\,K) giant planet orbiting the white dwarf within the forbidden zone (the region where planets are expected to be destroyed during the star's red giant phase). We constrain the source of the IR excess to an orbital separation of 0.1-2\,AU, marking the first discovery of a white dwarf planet candidate within this range of separations. WD~0310-688 is a young remnant of an A or late B-type star, and at just 10.4\,pc it is now the closest white dwarf with a known planet candidate. Future JWST observations could distinguish the two scenarios by either detecting or ruling out spectral features indicative of a planet atmosphere.  

\end{abstract}


\section{Introduction} \label{sec:intro}

White dwarfs represent the final evolutionary stage for most stars, including our Sun. Despite this common endpoint, the fate of their planetary systems remains poorly understood. Detection of exoplanets around white dwarfs have been challenging using traditional methods suitable for main-sequence stars \citep{Endl2018,2021ARA&A..59..291Z}. To date, only a few white dwarf exoplanets have been confirmed.

Due to the limited number of known white dwarf planets, we lack a comprehensive understanding of planetary evolution during the white dwarf phase. Increased detections could facilitate demographic studies and detailed characterization of such systems \citep{2005ApJ...633.1168D,veras2021planetary,2023A&A...675A.184L,2024AJ....167..257P}. This knowledge could help us ascertain whether planets can endure the death of their host stars and remain in orbit around the resultant white dwarf, as well as identify the conditions under which planets might be disrupted or destroyed during the red giant phase \citep{2002ApJ...572..556D,2010ApJ...722..725Z}. Furthermore, a deeper understanding of white dwarf planetary systems could reveal whether life can arise around dead stars \citep{2011ApJ...731L..31A,2012ApJ...757L..15F,2013AsBio..13..279B,2013MNRAS.432L..11L,2018ApJ...862...69K,2020ApJ...894L...6K,2023ApJ...945L..24B,2024arXiv240603189Z}.

The near-featureless {infrared} spectrum of white dwarfs is a valuable characteristic for exoplanet searches using the detection of infrared (IR) excess, which can indicate the presence of {sources colder than the white dwarf within the system} such as debris disks, late-type stellar companions, brown dwarfs, or exoplanets \citep{2005ApJ...632L.115K,2005ApJ...632L.119B,2008ApJ...674..431F,2012ApJ...760...26B,2020ApJ...898L..35S,Limbach22}. The IR excess technique enabled the discovery of the first brown dwarf-white dwarf system \citep{1987Natur.330..138Z,2022Natur.602..219C} and has since been employed to identify numerous similar systems \citep[e.g., ][]{2011MNRAS.417.1210G, 2015ApJ...806L...5X, 2019MNRAS.489.3990R, 2020MNRAS.498...12H, lai2021infrared}. Although the Spitzer Space Telescope was used in attempts to detect exoplanet-induced IR excess in white dwarf systems \citep{2004ApJS..154....1W, 2007ApJS..171..206M, 2008ApJ...681.1470F, 2010ApJ...708..411K}, no candidates were identified.

Fortunately, with the advent of the JWST, our ability to detect exoplanets around white dwarfs has drastically improved. JWST has multiple programs designed to detect new white dwarf exoplanets \citep{2021jwst.prop.1911M,2023jwst.prop.4403M,2023jwst.prop.3964M}. These search programs have resulted in the detection of multiple candidates \citep[][and {\it this work}]{2024arXiv240113153M} and follow-up observations to confirm planet candidates are underway \citep{2023jwst.prop.3621V,2024jwst.prop.6410C, 2024jwst.prop.4857M}. JWST's capabilities also extend to characterizing known white dwarf exoplanets \citep{2021jwst.prop.2358M,2021jwst.prop.2507V,2024jwst.prop.5204L,2024jwst.prop.6078B}.

Despite these impressive advancements, no known exoplanets (confirmed or candidate) exist around white dwarfs at mid-separations (0.1-2\,AU). Depending on the mass of the star, exoplanets beyond $\sim$2\,AU are expected to survive the final stages of stellar evolution intact \citep{Nordhaus2013MNRAS}. Discovering planets near or inward of this anticipated boundary illuminates our understanding of post-main-sequence exoplanet evolution and migration. The recent discovery of a volatile-rich gaseous debris disk from an evaporating giant planet \citep{2019Natur.576...61G} and the discovery of a gas giant exoplanet orbiting at only 0.02\,AU \citep{Vanderburg_2020} suggest that massive planets may find their way into tight orbits around white dwarfs, and the discovery of dusty disks {(generally hotter than giant planets but colder than the white dwarf)} around about 2-3\% of white dwarfs indicates that large rocky bodies may be torn apart by gravitational tides, producing a closely orbiting hot dust disk. The only other known white dwarf exoplanets and planet candidates orbit beyond the 2\,AU boundary where survival is expected \citep{1993ApJ...412L..33T,2003Sci...301..193S,2011ApJ...730L...9L,2021Natur.598..272B,2024arXiv240113153M}.

The detection of exoplanets around metal-polluted white dwarfs presents a unique opportunity to explore the intricate dynamics between stellar remnants and planetary systems. {Metal-rich white dwarfs are characterized by the presence of heavier elements in their atmospheres. This is surprising because metals should rapidly settle (on timescales as fast as days for hydrogen envelope white dwarfs \citealt{2006ApJ...646..474K,2006A&A...453.1051K,2007ApJ...663.1285J,2009A&A...498..517K})  below the white dwarf's surface due its strong gravitational fields, leaving only lightweight elements like hydrogen and helium in the atmosphere. 
Therefore, the presence of metals in white dwarfs atmosphere indicates that they are likely undergoing recent or active accretion. 
Studies in recent decades have concluded that the source of this accretion (also referred to as pollution) is usually minor planets, comets and asteroids from the progenitor system that survived red giant evolution \citep{1986ApJ...302..462A, 2003ApJ...584L..91J,2010ApJ...709..950K,2010ApJ...719..803D,2012ApJ...747..148D,2024RvMG...90..141V}. They likely moved close to the white dwarf due to gravitational interactions with other massive bodies, such as giant planets, in the system.} Small bodies, once destabilized, can migrate inward and eventually cross the white dwarf’s Roche limit, where they are tidally disrupted into dust and gas. This debris, rich in metals, then accretes onto the white dwarf, polluting its atmosphere. The chemical fingerprints of these accreted materials closely mirror the composition of terrestrial planets, such as Earth, providing strong evidence for this model \citep{2011ApJ...732...90M, 2019AJ....158..242X, 2023PSJ.....4..136T}.

Discovering white dwarf systems with planets and disks is crucial to validating the current theories on white dwarf accretion and planetary survival post-main sequence. In this paper, we present the detection of infrared excess at $<$2\,AU around the metal-polluted white dwarf WD 0310-688. Our observations, conducted as part of the MIRI Exoplanets Orbiting White dwarfs (MEOW) JWST survey program, are detailed in Section~\ref{sec:obs}. Section~\ref{sec:Analysis} covers our analysis of the MEOW data, focusing on the detection of infrared excess in the WD 0310-688 system. We use our MEOW data, modeling and archival data to constrain the source of infrared excess and its physical parameters. In Section~\ref{sec:discuss}, we discuss the implications of this detection and outline how future observations could confirm the source as an exoplanet. Finally, we summarize our findings in Section~\ref{sec:conclude}.

\begin{figure*}
\centering
\includegraphics[width=1\textwidth]{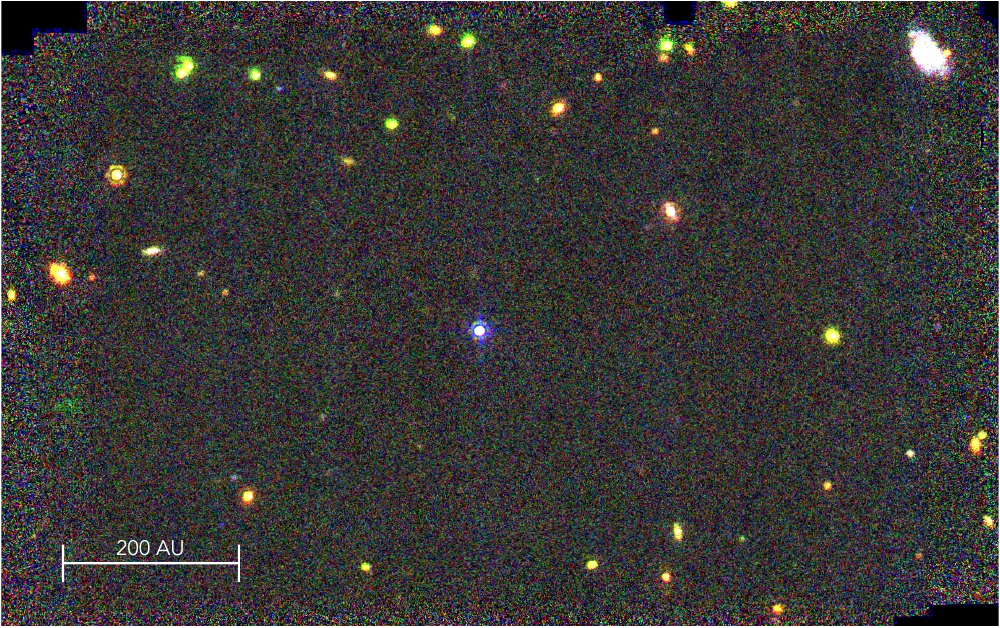}
\caption{False color image of WD 0310-688 (bluish-white star in the center) using three JWST/MIRI bands: F770W -- blue, F1800W -- green, and F2100W -- red. Other sources in the field include distant galaxies, background stars, and potentially a few point sources that could be wide-orbit exoplanets bound to WD 0310-688. Although not the main focus of this paper, we note the very faint source 80\,AU above the white dwarf, which is marginally detected at 18 and 21 $\mu m$, is consistent with a $T_{\rm eff}\sim$140\,K planet slightly less massive than Jupiter, however, at this separation the false positive rate is reasonably high. Confirming the any resolved sources as bound exoplanets would necessitate follow-up observations for common proper motion at a later epoch.}
\label{MIRIim}
\vspace{3mm}
\end{figure*}

\section{Observations} \label{sec:obs}

The MEOW survey is a JWST Cycle 2 Survey program (analogous to HST ``Snapshot" programs) to collect three-band MIRI imaging of $\sim$20 nearby white dwarfs in search of exoplanets. Utilizing a combination of infrared excess and direct imaging techniques, the MEOW survey is capable of detecting white dwarf exoplanets at all separations, from the Roche limit to the edge of the field of view ($\sim$1000\,AU; depending on the exact distance to the system).
MEOW is sensitive to extremely cold planets, with the ability to detect exoplanets with masses as low as that of Saturn and temperatures down to 120\,K for the youngest, nearest systems, and planets with temperatures as low as 175\,K, equivalent to a 2\,M$_{\rm Jup}$ at 3\,Gyr, out to a distance of 16\,pc. 

The MEOW survey is volume limited ($<17$\,pc). It includes solitary white dwarfs and white dwarfs with faint companions (main sequence stars later than M4V or binary white dwarfs). White dwarfs with bright main sequence companions (e.g., Sirius B, Procyon B, 40 Eridani B, etc) are excluded as a different imaging configuration would be optimal for detecting exoplanets in those systems.

Imaging is taken in three MIRI bands: F770W, F1800W and F2100W (7.7, 18.0 and 21.0 $\mu$m, respectively). For the F770W filter, our total exposure time is 55.5\,sec (5 groups/int, 1 int/exp), which is sufficient to achieve an SNR$>$200 on all targets in our sample while still avoiding saturation on the nearest white dwarfs. The SNR in 7.7\,$\mu$m band is dominated by photon noise from the target star; however, our absolute photometric precision is limited to 2\% by the absolute flux calibration of the MIRI imager\footnote{\url{jwst-docs.stsci.edu/jwst-data-calibration-considerations/jwst-data-absolute-flux-calibration}} \citep{2022arXiv220406500G}. For the F1800W and F2100W filters, our exposure times are 277.5\,sec and 710.4\,sec, respectively (12 groups/int, 2 or 5 int/exp), which results in an SNR between 10 and 125 on all targets in our sample. Depending on the target brightness, the noise in the 18 and 21\,$\mu$m bands can be dominated by either the thermal background noise or the 2\% absolute photometric precision.

For all imaging, we used the fast readout mode and a four point cycling dither (starting point 1). We chose a 4-point cycling dither to correct for bad pixels and remove background noise. We used the full MIRI imaging array, which has a 74" $\times$ 113" useable field of view, and a detector plate scale of 0.11"/pixel \citep{2015PASP..127..612B}. 

The observations of WD 0310-688 began Sep 19, 2023 22:50:31 UT and ended at Sep 19, 2023 23:53:06, about an hour later\footnote{Of note, this star was the very first JWST survey observation ever conducted, for any survey program.}.
The false color image constructed from the MIRI imaging in the three bands is shown in Figure \ref{MIRIim}. 

{Some initial findings from the MEOW survey are included in this Letter as they relate to the infrared excess detected around WD~0310-688. However, the majority of the survey's results and analysis will be detailed in a future manuscript. This Letter primarily focuses on reporting the discovery of a planet candidate around WD~0310-688.}

\section{Analysis} \label{sec:Analysis}

\subsection{The White Dwarf WD 0310-688}\label{WDparams}
WD 0310-688 is a solitary white dwarf located at a distance of 10.4\,pc. With $V$ and $G$-band magnitudes of 11.4~mag, it is the 5$^{th}$ brightest white dwarf at visible wavelengths and the single brightest solitary white dwarf in the sky.\footnote{The five brightest white dwarfs in the optical are Sirius~B ($V=8.4$), 40~Eri~B ($V=9.5$), Procyon~B ($V=10.9$), CD-38~10980 ($V=11.0$), and WD~0310-688 ($V=11.4$) \citep{Holberg2008, 2021MNRAS.508.3877G}.}

\subsubsection{Stellar Parameters}
The spectroscopic \citep{2017ApJ...848...11B} and photometric \citep{2021MNRAS.508.3877G} determinations of WD 0310-688's atmospheric parameters differ slightly, so we calculate the ages based on both sets of parameters. This white dwarf is in a regime where a small difference in its mass can significantly affect its main-sequence lifetime. We calculated the white dwarf parameters using \texttt{wdwarfdate}, and the resulting parameters are listed in Table \ref{tab:parameters}. These parameters suggest that the host is a remnant of an A or late B star. We base our calculations on photometric white dwarf parameters because the spectroscopic parameters yield a slightly low white dwarf flux that is in tension with Gaia photometry, thus favoring the photometric solution. The photometric parameters give a white dwarf mass of $0.659 \pm 0.012$ M$_\odot$ and cooling age of $194 \pm 13$\,Myr.

\begin{table}
\centering
\caption{WD 0310-688 Spectroscopic \& Photometric Derived Parameters}
\vspace{-3mm}
\label{tab:parameters}
\begin{tabular}{c|c|c}
Parameter & Spectroscopic$^a$ & Photometric$^b$ \\ 
\hline 
T$_{\rm eff}$ (K) & $16,246 \pm 237$ & $15,865 \pm 263$ \\
log\,$g$ & $8.14 \pm 0.04$ & $8.076 \pm 0.020$ \\
Initial mass (M$_\odot$) & $2.3_{-0.5}^{+0.4}$ & $1.9 \pm 0.4$ \\
MS lifetime (Gyr) & $0.9_{-0.3}^{+0.6}$ & $1.4_{-0.5}^{+1.4}$ \\
Total age (Gyr) & $1.1_{-0.3}^{+0.6}$ & $1.6_{-0.5}^{+1.4}$ \\ 
\end{tabular}
\\
\vspace{1mm}
{\footnotesize $^a$Spectroscopic Teff and log(g) from  \citealt{2017ApJ...848...11B}; $^b$Photometric Teff and log(g) from \citealt{2021MNRAS.508.3877G}.}
\end{table}

\begin{figure*}
\centering
\includegraphics[width=0.88\textwidth]{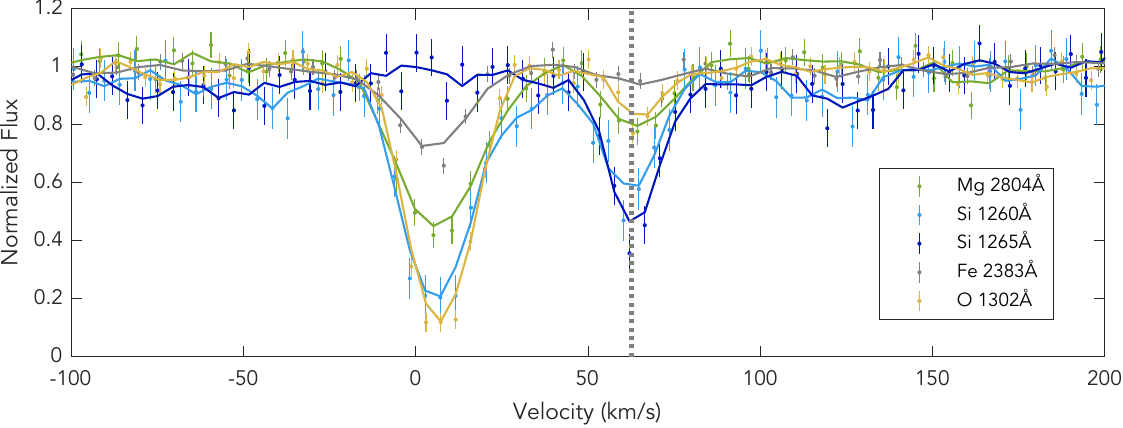}
\caption{In a reanalysis of an archival HST/STIS spectrum of WD 0310-688, we detect atomic lines of four metals in the white dwarf's photosphere. This suggests accretion of metal-rich material on the white dwarf, and we reclassify it as a DAZ due to the presence of the metal lines. WD~0310-688 was previously classified as a DA spectral type.}
\label{STISspectrum}
\end{figure*}

\subsubsection{Metal Lines \& Accretion}\label{pollution}
WD 0310-688 is classified in the literature as a hydrogen-atmosphere (DA) white dwarf with no metal lines, and has been observed as part of ground based observation campaigns \citep{2009A&A...505..441K}. However, in {a previously-unpublished archival UV} HST/STIS E140M+E230M spectrum of this white dwarf \citep{2015hst..prop14076G}, which we retrieved from the HASP database \citep{2024cos..rept....1D}, we find O, Si, Mg and Fe absorption in the spectrum near the photospheric radial velocity of WD 0310-688 (see Figure \ref{STISspectrum}). This suggests accretion of metal-rich material on the white dwarf, and we reclassify it as a DAZ. We also note the presence of the Si\,II\,1265\,${\rm \AA}$ line indicating that the UV absorption lines are not from the interstellar medium \citep{2014A&A...566A..34K} and confirming that there is pollution in the white dwarfs photopshere. Accretion has implication for the disk and exoplanet hypotheses, which we discuss more in Section \ref{sec:discuss}, so we will further constrain this accretion here.

We construct a model of WD 0310-688 using an updated version of the model atmosphere code described by \cite{blouin2018a,blouin2018b} and references therein. The model atmosphere code has been modified to include all relevant spectral lines and molecular bands up to $30\,\mu$m as described by \cite{Limbach22}. 

We calculate a log N(Si)/N(H) = -8.1$\pm$0.2 from the STIS spectrum by fitting three Si II lines (at 1260.42, 1264.73 and 1265.02\,${\rm \AA}$). For this white dwarf with a temperature of 15,865\,K and log(g) of 8.076, the diffusion timescale of Si is 1.2$\times10^5$ seconds (1.4 days; \citealt{Dufour2017}). The short diffusion timescales for this white dwarf are due to its high temperature, indicating that it must have been accreting in 2016 during the STIS observations. The thickness of the superficial convection zone is 10$^{-16.6}$ of the total mass \citep{Dufour2017}. By dividing the mass of Si mixed in the convection zone by the sinking timescale, we obtain a steady-state accretion rate of Si of 5.3$\times10^4$\,g/s. 
{Relative to Si, we find that the other metal abundances, listed in Table \ref{tab:parameters2}, are consistent with the bulk Earth \citep{1995E&PSL.134..515A} once accounting for diffusion timescales. For Fe, the detection is marginal so it is not given in the table, but if we assume a log Fe/Si that corresponds to bulk Earth, we get a synthetic spectrum consistent with the HST spectrum.}

Assuming the accreted material has a bulk Earth composition {and based on the steady-state accretion rate of Si}, the total accretion rate is 3.3$\times10^5$\,g/s. This is among the lowest accretion rates observed in polluted white dwarfs \citep{2022MNRAS.510.1059B}. A summary of all the white dwarf's parameters from this work and the literature is provided in Table \ref{tab:parameters2}.

\begin{table}[h]
\centering
\caption{Summary of WD 0310-688 Parameters}
\label{tab:parameters2}
\begin{tabular}{c|c|c}
Parameter & Value & Ref\\ 
\hline 
Name & WD 0310-688 &\\
Alt. Name & CPD-69 177 &\\
RA (ep=J2000) &	03 10 31.0195960 & 1\\
DEC (ep=J2000) &-68 36 03.380768 & 1\\
RA PM (mas/yr) & 39.668$\pm$0.036  &1\\
DEC PM (mas/yr) & -103.237$\pm$0.032 & 1\\
RV (km/s) & 62.6$\pm$0.4 & 2\\
Parallaxes (mas) & 96.1834$\pm$0.0289 & 1\\
Distance (pc) & 10.397$\pm$0.003 & 1\\
Spectral type & DAZ & 3\\
U (mag) & 10.757 & 5\\
G (mag) & 11.409583 & 1\\
K (mag) & 11.861 & 6\\
T$_{\rm eff}$ (K) & $15,865 \pm 263$ &3,4\\
log\,$g$ & $8.076 \pm 0.020$ &3,4 \\
Initial mass (M$_\odot$) & $1.9 \pm 0.4$ &3,4\\
WD Mass (M$_\odot$) & $0.659 \pm 0.012$ &3,4\\
MS lifetime (Gyr) & $1.4_{-0.5}^{+1.4}$ & 3,4\\
WD age (Gyr) & 0.194$\pm$0.013 & 3,4\\
Total age (Gyr) & $1.6_{-0.5}^{+1.4}$ &3\\ 
Accretion rate (g/s) & 3.3$\times10^5$ & 3\\
log\,N(Si)/N(H) & -8.1$\pm$0.2 &3\\
log\,N(O)/N(H) & -7.2$\pm$0.2&3\\
log\,N(Mg)/N(H) & -8.0$\pm$0.2&3
\end{tabular}
\\
\vspace{1mm}
{\footnotesize Refs: 1.~\citealt{2020yCat.1350....0G}, 2.~\citealt{2020A&A...638A.131N}, 3.~{\it this work}, 4.~\citealt{2021MNRAS.508.3877G}, 5.~\citealt{2010MNRAS.403.1949K}, 6.~\citealt{2003yCat.2246....0C}}
\vspace{1mm}
\end{table}

\subsection{MEOW Data Analysis}
{

\subsubsection{Resolved Sources in the Image}
}

{Prior to conducting a detailed analysis, we visually inspected the color image for resolved sources that could be bound companions to the white dwarf. The purpose of this brief search is to note the presence of possible wide-orbit companions that could serve as perturbers (via planet-planet scattering; see section \ref{sec:discuss}), which is relevant to the IR excess detection presented within this manuscript.}

We identify several resolved sources in Figure \ref{MIRIim} that exhibit colors and fluxes consistent with exoplanets. One of these is relatively close, about 80\,AU from the white dwarf. This source (which is  not the primary focus of this manuscript), is only detected with a signal-to-noise of a few, but is consistent with a $\sim$140\,K planet (which given the system's age, would likely be sub-Jovian in mass; \citealt{2019A&A...623A..85L}). However, at this angular separation the false positive rate is reasonably high in our survey. There are also many other sources at larger angular separations from the white dwarf with photometry consistent with warmer, more massive exoplanet models. We expect that most or all of these point sources {at large angular separations} are background objects, but additional epochs of MIRI imaging could show whether any share common proper motion with the white dwarf, indicating they are bound exoplanets. This information will be crucial for understanding exoplanet occurrence rates around white dwarfs and the systems' dynamical histories. However, we defer detailed consideration of the resolved point sources around WD 0310-688 until follow-up observations are conducted.

\subsubsection{Photometry and IR Excess}
In this subsection, we describe how we (1) reduce the data using a custom pipeline, (2) perform aperture photometry and report the measured flux values, (3) quantify the measurement errors, (4) compute the infrared excess from the photometry, and (5) provide checks to verify that the measured infrared excess is real and not attributable to any emission from the white dwarf itself.

\begin{center}
    \vspace{-1mm}
    {\it Step 1: Data Reduction}
    \vspace{-2mm}
\end{center}
The MIRI imagery reduced automatically by the JWST pipeline and available in MAST suffers from a non-uniform flatfield subtraction at the reddest wavelengths.  {To address this, we reprocessed the data using a custom software package, \texttt{MEOW}, that is available on GitHub\footnote{\url{https://github.com/kevin218/MEOW}}. Version 1.0 of \texttt{MEOW} was used for this reduction.  We used JWST pipeline version 1.15.1 and CRDS jwst$\_$1225.pmap.} The background subtraction code is based on a STScI JWebbinar demo\footnote{\url{https://github.com/spacetelescope/jwebbinar_prep/blob/jwebbinar31/jwebbinar31/miri/Pipeline_demo_subtract_imager_background-platform.ipynb}} that produces well flat-fielded Stage 2 data with a custom background subtraction using the multiple dithers on each source. 

\begin{center}
    \vspace{-1mm}
    {\it Step 2: Aperture Photometry}
    \vspace{-2mm}
\end{center}
Using the reprocessed Stage 2 data, we then conducted aperture photometry on the white dwarfs in the MEOW dataset. To conduct aperture photometry we used the Python \texttt{photutils} package and the aperture correction values provided in the JWST CRDS\footnote{\url{https://jwst-crds.stsci.edu/}; file version \path{jwst_miri_apcorr_0017.rmap}} and color corrections table provided by the STScI HelpDesk\footnote{\url{https://stsci.box.com/shared/static/47xxtu2yq21tdfioey7q8dsd4ly8uwki.dat}; \path{color_corrections_miri.dat}}. For the aperture sizes, we used the values corresponding to the full array (as we read out the full MIRI subarray) and the 80\% energy encircled aperture sizes (the largest apertures with correction values available) on the target PSF.

The resulting measured flux values are given in Table \ref{WDflux}. In each band we measure the flux of the white dwarf at each of the four dithers separately and then report the white dwarf's median flux value, using the scatter in the flux measurements divided by the square-root of the number of dithers to compute the error of the measured flux value.

\begin{table}[b]
\centering
\caption{Measured WD flux and IR excesses.}
\label{WDflux}
\begin{tabular}{c|c|c}
Band & Flux & IR Excess ($F_p/F_*$)\\ 
\hline 
F770W & 1063.5$\pm$2.9\,$\mu$Jy & -0.9$\pm$2\%\\ 
F1800W & 230.4$\pm$0.8\,$\mu$Jy  & 16.6$\pm$2\% \\ 
F2100W & 179.1$\pm$0.7\,$\mu$Jy  & 21.3$\pm$2\% 
\end{tabular}
\\
\vspace{1mm}
\end{table}

\begin{center}
    \vspace{-1mm}
    {\it Step 3: Error Sources}
    \vspace{-2mm}
\end{center}
The errors in the infrared excess measurement of white dwarfs in the MEOW sample arise from two primary sources: (1) the precision (signal-to-noise) of the measurement, which is primarily limited by photon noise (listed in the middle column in Table \ref{WDflux} for WD 0310-688), and (2) the accuracy with which our white dwarf models predict the star's flux. For the MEOW data, the limiting factor can either be photon noise or the accuracy of the JWST photometry and our models, depending on the bandpass and the brightness of the white dwarf. Notably, WD 0310-688, one of the brightest white dwarfs, is consistently limited by model/photometric accuracy across all bands.

\begin{figure}
\begin{center}
\vspace{-2mm}
\leavevmode
\begin{minipage}[c]{0.49\textwidth}
\includegraphics[width=0.95\textwidth]{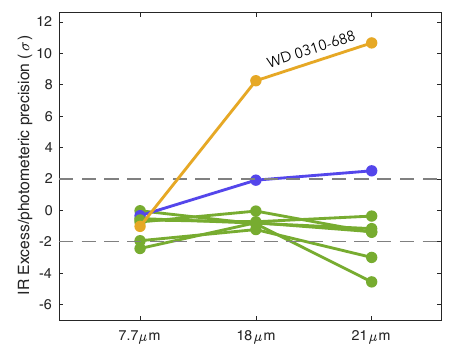}
\end{minipage}\hfill
\begin{minipage}[c]{0.47\textwidth}
\caption{The measured IR excess or deficit divided by the precision of our measurement (y-axis) in all three spectral bands (x-axis) for a subset of MEOW white dwarfs. The orange line represents WD 0310-688, which shows a highly significant IR excess. The purple line corresponds to another white dwarf with an excess possibly due to an exoplanet but detected with less significance than WD 0310-688. The green lines indicate no significant excess, aligning generally with our models, although they are systematically low by 1-2\% at 7.7$\mu$m and 21$\mu$m. The white dwarf that is notably low at 21$\mu$m is likely affected by a nearby bright, red galaxy contaminating the background flux measurement.}
\label{MEOWexcess}
\end{minipage}
\end{center}
\end{figure}

\begin{center}
    \vspace{-1mm}
    {\it Step 4: Infrared Excess}
    \vspace{-2mm}
\end{center}
We compute the infrared excess, defined as the measured flux of the white dwarf relative to the expected (modeled) flux values. We build a model for each white dwarf using the same modeling process described in Section \ref{WDparams} for WD 0310-688. This analysis and IR excess calculation was conducted for all white dwarfs in the MEOW sample. Figure \ref{MEOWexcess} illustrates the infrared excess (or deficit) divided by the precision of our measurement for a subset of the white dwarfs in the MEOW sample. Only white dwarfs without a large ($>$100 MG) magnetic field and those amenable to aperture photometry\footnote{For many MEOW targets, PSF fitting is necessary due to nearby sources contaminating the background.} are included in this plot. {The white dwarfs included in the plot are WD~0839-327, WD~0840-136, WD~1309+853, WD~1820+609, WD~1756+827, WD~1019+637, WD~0752-676, WD~0821-669 and WD~0310-688.} This measurement precision accounts for both error sources (1) and (2). In this plot, we take error (2) to be 2\%, the absolute photometric precision of MIRI as discussed in Section \ref{sec:obs}. 

In Figure \ref{MEOWexcess}, the green lines representing white dwarfs without notable IR excess are systematically low by 1-2\%. To address this, we apply a correction factor to the measured IR excesses where we have a sufficiently large set of white dwarfs with a signal-to-noise ratio greater than 50 that can be used for calibration. At 7.7$\mu$m, almost all white dwarfs meet this criterion and at 18$\mu$m only a couple of the brightest white dwarfs in our sample do. The derived correction factors are: 1.14\% at 7.7$\mu$m and 0.1\% at 18$\mu$m.  At 21$\mu$m, there are no comparatively bright white dwarfs that provide a good reference, so we do not apply an additional correction factor at 21$\mu$m. 

The IR excesses for WD0310-688 (orange line), including the correction factors, are -0.9\%, 16.6\%, and 21.3\% at 7.7$\mu$m, 18$\mu$m, and 21$\mu$m bands, respectively (as shown in Figure \ref{IRexcess} and listed in Table \ref{WDflux}). If the source of the excess is a planet, the infrared excess corresponds to the flux ratio of the planet to the star ($F_p/F_*$).  
We calculate the significance of the IR excess detected in the WD 0310-688 system to be 8.3$\sigma$ at 18$\mu$m and 10.7$\sigma$ at 21$\mu$m bands. No significant excess is detected at 7.7$\mu$m.

{The purple line in Figure \ref{MEOWexcess} indicates a marginal detection of IR excess for WD~1309+853. This white dwarf has a relatively weak magnetic field of 5 MG, which could potentially account for the discrepancies between the photometry and the model. None of the other white dwarfs presented in this plot are known to have magnetic fields. A full analysis and discussion of the IR excesses within the MEOW sample is deferred to a later manuscript.} 

\begin{center}
    \vspace{-1mm}
    {\it Step 5: Verifying the Result}
    \vspace{-1mm}
\end{center}
To further validate the robustness of the IR excess measurement for WD 0310-688, we conducted two additional tests. First, we compared the fluxes using the Stage 3 data available in MAST for our target, finding the flux values agreed with our custom reduction results to within 2\% in all bands (some deviation is expected due to non-uniform flatfield in the Stage 3 data).

Second, we generated a range of white dwarf models consistent with the $\pm1\sigma$ uncertainties derived from the optical photometry and spectroscopy described in section \ref{WDparams}. Our analysis revealed that varying the white dwarf model does not produce an IR excess that increases with wavelength, as observed with MIRI. Instead, adjusting the white dwarf parameters results in a near-uniform shift in the IR excess across all MIRI wavelengths, owing to the fact that MIRI photometry samples the Rayleigh-Jeans tail of the white dwarf's SED.

\begin{figure*}
\centering
\includegraphics[width=0.85\textwidth]{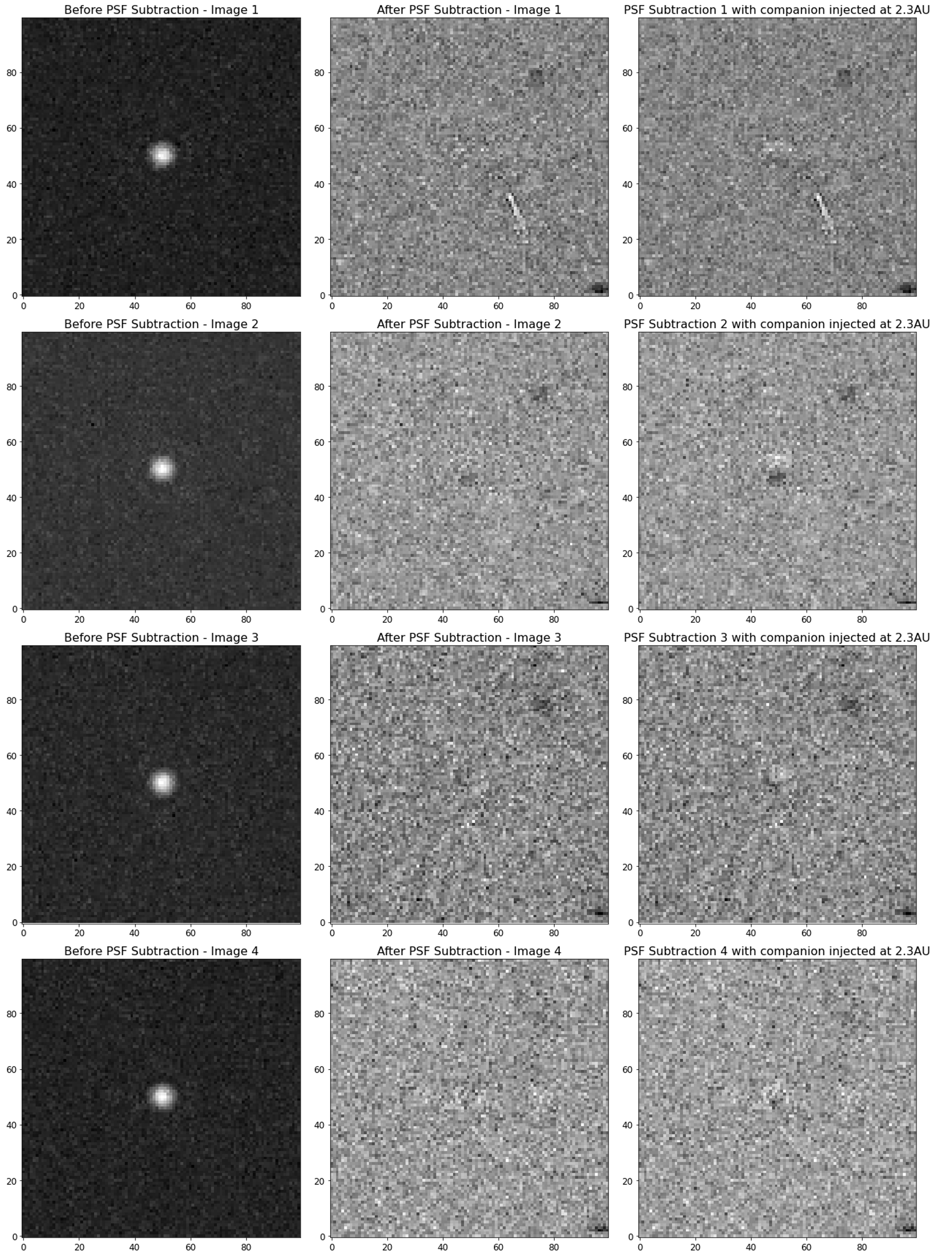}
\caption{{\bf Left:} WD 0310-688 at 21\,$\mu m$ in each of the four dithers, {\bf Center:} PSF Subtraction at each of the four dithers. Note that the cosmic ray in the first dither and the negative source at a $\sim$30 pixel separation is in the reference star image, not the science image. {\bf Right:} PSF subtraction after a companion that is 21\% the brightness of the star was injected at 2.3\,AU. This injected source is retrieved with a statistical significance of 7.0\,$\sigma$ (at each dither the significance of the detections are 3.2\,$\sigma$, 4.2\,$\sigma$, 3.7\,$\sigma$ and 2.7\,$\sigma$). Injected sources are successfully retrieved at $>$4\,$\sigma$ when injected at separations of $>$2\,AU and therefore we constrain the detected IR excess source to be within 2\,AU of the host.}
\label{PSF}
\end{figure*}

\subsubsection{PSF Subtraction}
In this section, we use PSF subtraction to constrain the maximum orbital separation of the IR excess source and determine the background source false positive probability.
In an attempt to directly image the source emitting the infrared excess, we conducted a PSF subtraction using our custom flat-field corrected 21\,$\mu m$ Stage 2 data.  We used four images of WD 0310-688 for the science frames (see Figure \ref{PSF}) and four reference images from the white dwarf  {WD~0839-327 (G = 11.82\,mag), the white dwarf most similar to WD~0310-688 in our MEOW sample, as our reference PSF. }
We used the python package \texttt{VIP} \citep{VIP} to crop the images to a size of 100px (1 arcsec), correct NaN values, and align the PSFs to a common center. 
We then scaled the reference PSF to the science PSF and performed a PSF subtraction (results are shown in Figure \ref{PSF}, center column). We find no notable residuals in the subtraction suggesting the source of infrared excess is unresolved from the star. 

We then injected companions with the same flux as the detected IR excess at various separations from the target. We find we are able to detect the injected companions after PSF subtraction with a confidence of $>$4\,$\sigma$ when the companion is injected at separations of $>$2\,AU, and therefore we constrain the source to be within 2\,AU of the host.

The PSF subtraction constrains the maximum separation of the infrared excess source to $<$2\,AU, or $<0.19$". Using this maximum separation, we calculated the probability that our infrared excess is due to a background object rather than something in the WD 0310-688 system. The MIRI field of view is 112.6" $\times$ 73.5", which corresponds to a false positive rate of 1/73000 if there was one false positive in the FOV and we observed only one white dwarf. However, MEOW images typically contain four sources within the field of view that are consistent with exoplanet SEDs, and the MEOW survey has observed 17 targets. Assuming the four planet-like sources per field are false positives (though some may actually be planets), we estimate the probability that the IR excess we have detected is a background source to be 1/73000 times 4 FPs/field and 17 targets, giving a false positive probability of $<$0.1\%. {We note that none of the MEOW targets with reliable photometry (no contamination from nearby sources) exhibit an IR excess, except for those with known magnetic fields.} We also note that the WD~0310-688 field is less dense than most of the others within the MEOW survey, making this estimate likely conservative. Due to the very low false positive probability, we conclude that the IR excess is almost certainly associated with the white dwarf system.

\subsubsection{Blackbody Retrieval}

Having ruled out background sources as a plausible explanation for the infrared excess, we conclude that the IR excesses are likely due to a cold dust disk or an exoplanet. We now explore models that can explain the measured excess.

\begin{figure*}
\centering
\includegraphics[width=0.88\textwidth]{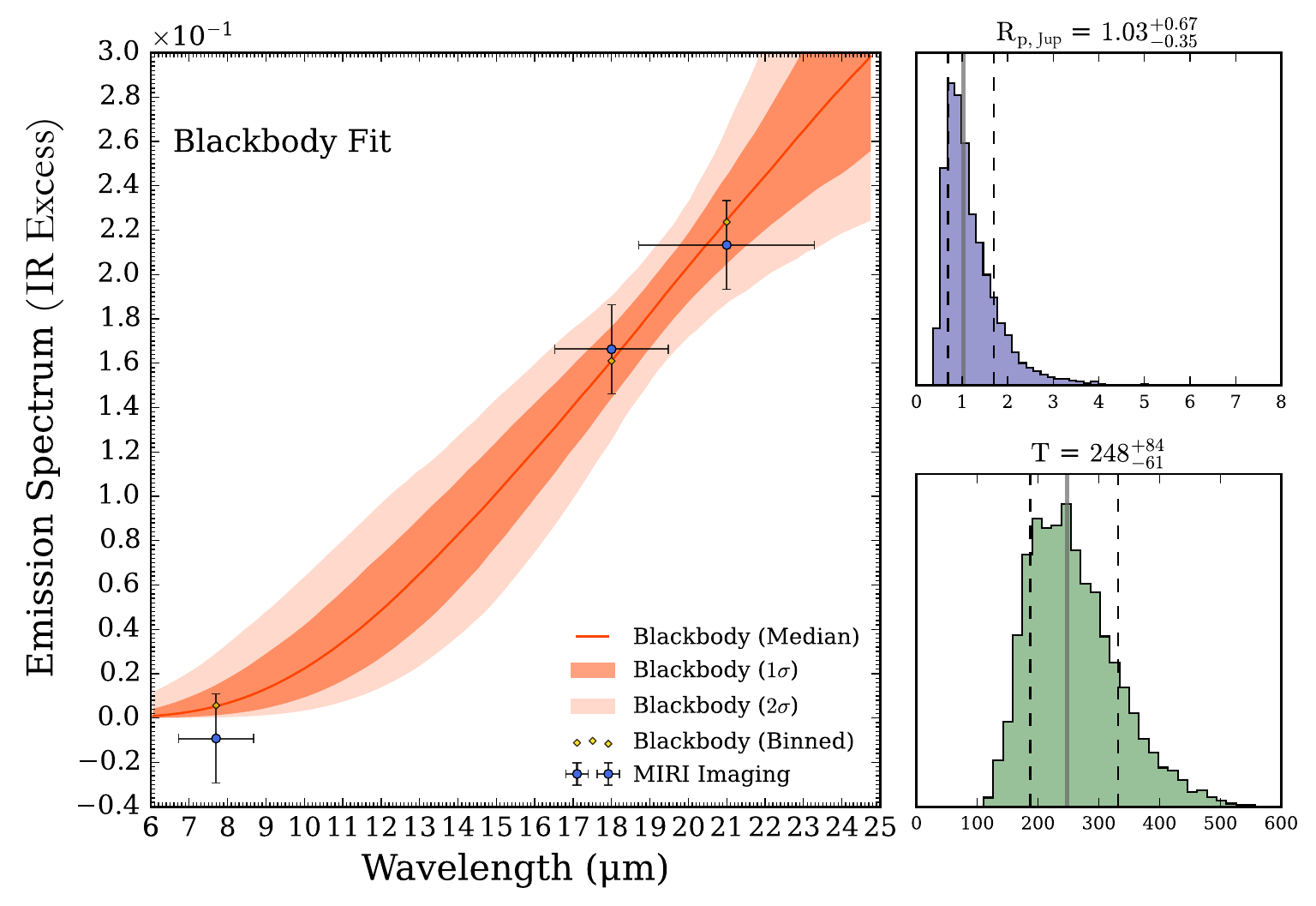}
\caption{Measured wavelength-dependent infrared excess (i.e., the planet-to-star flux ratio, if the source of the infrared excess is a planet) from three photometric bands collected with JWST MIRI (F770W, F1800W and F2100W). The infrared excess is consistent with a blackbody of temperature 248$^{+84}_{-61}$\,K and an emitting area of 1.03$^{+0.67}_{-0.35}$\,R$_{\rm Jup}$. Using the derived system age, the planet's temperature, and planet cooling curves, we estimate a planet mass of $3.0^{+5.5}_{-1.9}$M$_{\rm Jup}$.}
\label{IRexcess}
\end{figure*}

Because we only have three photometric data points, the data does not justify a model more complex than a blackbody. The advantage of a blackbody model is that it is agnostic as to whether the source is a disk or an exoplanet, as both are, to first order, blackbodies in the mid infrared.

We conducted a blackbody fit using the measured infrared excess. In Figure \ref{IRexcess}, we plot the measured excess alongside the temperature and size constraints from our blackbody fit. The fit to the photometric measurements yield a T = 248$^{+84}_{-61}$\,K blackbody with an emitting area of 1.03$^{+0.67}_{-0.35}$\,R$_{\rm Jup}$. 

We note that the size of the emitting body is conspicuously close to that of a giant planet, but we do not disregard the possibility that a cold dust disk could also have an emitting area coincidentally similar to that of a gas giant planet. However, this would constitute a very unusual cold dust disk, as discussed in Section \ref{sec:discuss}.
We attempted to run more complex atmospheric retrievals on the data, but found that this resulted in no additional meaningful constraints on the source. 

\subsubsection{Minimum Orbital Separation}

We calculate the minimum orbital separation, $a$, of the source using the fitted blackbody temperature as the equilibrium temperature, $T_{eq}$. This calculation assumes that the measured brightness temperature is equal to the blackbody temperature, which is a reasonable assumption in the mid infrared, and that there is perfect heat redistribution. The orbital separation is given by
\begin{equation}
 a = R_*(1-\alpha)^\frac{1}{2}\left(\frac{T_{eff}}{T_{eq}}\right)^2
\end{equation}
where $R_*$ is the radius of the star, $\alpha$ is the albedo of the planet, and $T_{eff}$ is the effective temperature of the white dwarf. This results in a minimum orbital separation of $0.18^{+0.14}_{-0.08}$\,AU (with $\alpha = 0.4$; a typical infrared albedo for a highly reflective gas giant). The 1$\sigma$ lower bound gives a minimum orbital separation of 0.10\,AU.  Orbital separations $<$0.1\,AU would result in higher brightness temperatures that are inconsistent with the fit. This constraint assumes a circular orbit. However, if the planet has a highly eccentric orbit (and is migrating), it may pass much closer to the white dwarf, becoming significantly hotter during parts of its orbit.
If the source is a optically thin disk (e.g., zodi-like) rather than a planet, the same orbital constraint ($0.18^{+0.14}_{-0.08}$\,AU) holds. The exception would be if the source is a planet and exhibits an exceptionally high albedo (e.g., $\alpha > 0.4$), which could potentially allow for slightly closer separations.

\subsubsection{Mass Estimation}\label{sec:mass}
If the source is an exoplanet, we can derive a mass estimate using exoplanet evolutionary models. 
Using the age of the system and the planet's temperature, along with the 1\,$\sigma$ uncertainties in these two parameters, we constrain the planet's mass to $3.0^{+5.5}_{-1.9}$\,M$_{\rm Jup}$ \citep{Marley2021}, assuming no heat comes from irradiation.
If the planet is out at 1-2\,AU, the stellar irradiation has essentially no effect on the inferred mass because it would be dominated by the internal heat of the planet, so at those distances the evolutionary masses apply, as previous studies suggest modest irradiation does not change the cooling very much \citep{2007ApJ...659.1661F}.

However, at closer separations, irradiation from the white dwarf could change the evolutionary inferred mass. If the planet orbits close to the white dwarf ($\sim$0.1\,AU), its flux must be dominated by re-radiated stellar radiation. In this scenario, the mass could be much lower. We can still place a lower limit on the mass based on the fact that we need the planet to be sufficiently large in radius. We estimate the minimum mass based on the minimum planet radius from our blackbody fit within the 1$\sigma$ constraint, which is 0.68\,R$_{\rm Jup}$. Using the \cite{2019A&A...623A..85L} models, at an age of 1.1\,Gyr, we estimate a minimum mass of 0.2\,M$_{\rm Jup}$ (slightly less massive than Saturn). However, we note that empirically, there are Jupiter-sized objects with masses as low as a few Earth masses, such as the Kepler 51 super puffs \citep{2014ApJ...783...53M,2020AJ....159...57L}.

\subsection{Other Constraints on Companions}

\subsubsection{Astrometry}

If WD~0310-688 is indeed orbited by a giant planet, we can place further constraints on its properties using astrometry. The most precise existing astrometric observations for this target are from the \textit{Gaia} mission \citep{Gaia}; however the \textit{Gaia} epoch astrometry is not yet available, precluding any straightforward search for companions. Nevertheless, it is possible to extract some limits on companion properties from the published \textit{Gaia}~DR3 astrometric solution.

The \textit{Gaia} Renormalized Unit Weight Error (RUWE) parameterizes the significance excess noise of the astrometric data. Conventionally, a RUWE above $>$1.4 is interpreted as significant evidence in favour of astrometric variability \citep{Lindegren18}. In \textit{Gaia}~DR3 WD~0310-688 has a RUWE of only 1.13, which suggests there is no strong evidence for orbital motion in the underlying epoch astrometry.

We then attempt to convert the RUWE into a set of upper limits on companion masses. We follow the method of \citep{Belokurov20, Korol22} to invert the RUWE into the normalized astrometric perturbation, $\delta\theta$. As the observed RUWE is not significant, we employ RUWE $<1.4$ as a conservative limit on the astrometric noise, which then gives an assumed limit of $\delta\theta<0.19$~mas. For orbital periods shorter than or approximately equal to the 1038~d observational baseline of \textit{Gaia}~DR3 \citep{Lindegren21}, we can assume that this approximates the root mean square of the astrometric reflex amplitude of the theorised orbit. This allows us to assume the conventional astrometric amplitude relation:

\begin{equation}
    \alpha=\frac{m}{M_*}\frac{a}{D}
    \:,
\end{equation}

where $\alpha$ is the astrometric amplitude in arcseconds, $m$ is the companion mass, $M_*$ is the stellar mass, $a$ is the semi-major axis in AU, and $D$ is the distance in parsecs \citep{Sozzetti05}. As $M_*$ and $D$ are known for WD~0310-688, we can derive an upper limit on allowed companion masses as a function of semi-major axis within $\lessapprox$2~AU.

More stringent constraints can be made on companions with wider orbits using \textit{Hipparcos-Gaia} astrometry \citep{Brandt2018, Kervella2019}. WD~0310-688 was one of only 20 white dwarfs bright enough to be successfully observed with \textit{Hipparcos} \citep[HIP~14754;][]{Vauclair1997}, and when combined with the more recent \textit{Gaia} astrometry, the 25~year time baseline allows for strong constraints on longer-period planets. In the \textit{Hipparcos-Gaia} Catalog of Accelerations \citep[HGCA;][]{Brandt2021} there is no significant evidence for an astrometric acceleration; the difference between the \textit{Gaia} proper motion and the averaged \textit{Hipparcos-Gaia} proper motion in right ascension and declination is $\Delta\mu=$~($-0.017\pm0.075$, $-0.012\pm0.080$~mas~yr$^{-1}$) respectively, equivalent to $\Delta v=$~($-0.8\pm3.7$, $-0.6\pm3.9$~m~s$^{-1}$) in physical units \citep[][equations 9, 10]{Venner2021}. We can thus set a strict 3$\sigma$ upper limit of $<$11~m~s$^{-1}$ on the \textit{Hipparcos-Gaia} tangential velocity anomaly.

\begin{figure}
  \begin{center}
    \vspace{-2mm}
      \leavevmode
        \begin{minipage}[c]{0.53\textwidth}
\includegraphics[width=0.95\textwidth]{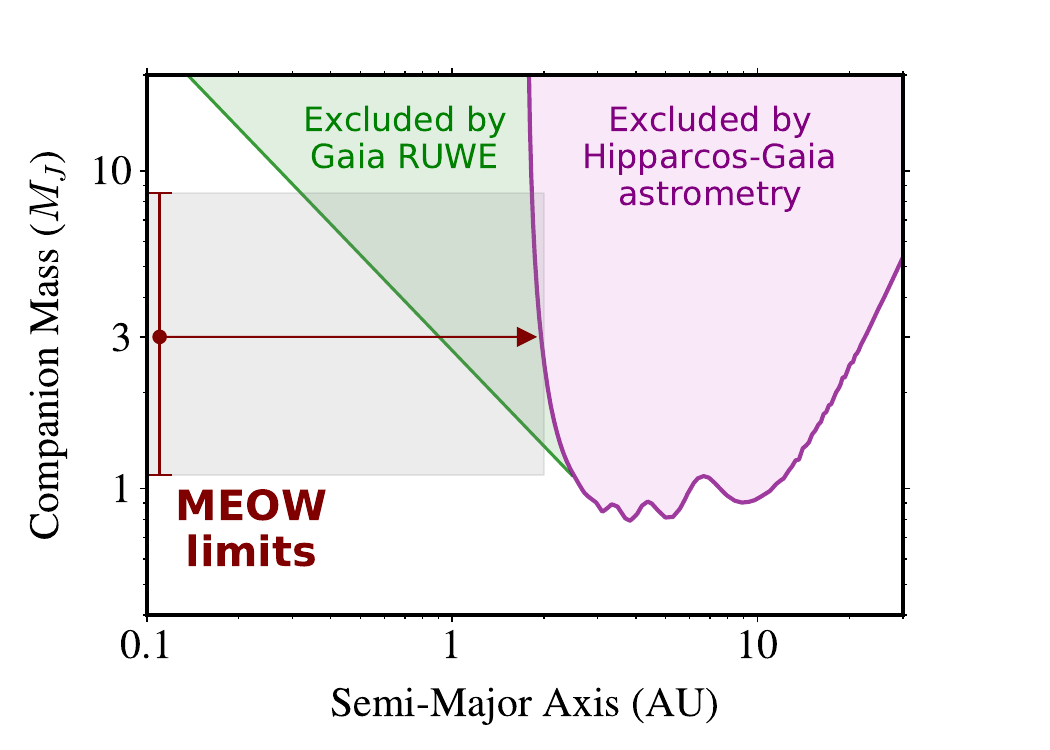}
  \end{minipage}\hfill
  \begin{minipage}[c]{0.47\textwidth}
\caption{Constraints on the planets orbiting WD~0310-688 from astrometry. We overplot the planetary parameters implied bu the infrared excess, assuming that the upper limit in projected separation translates to semi-major axis. The absence of a large \textit{Gaia}~DR3 RUWE helps to exclude the more distant, massive edge of the parameter space, but within $a<1$~AU the sensitivity of astrometry is low. At larger distances ($2.5-10$~AU), planets with as low mass as 1~$M_J$ can be ruled ruled out due to the absence of a significant \textit{Hipparcos-Gaia} acceleration.
} \label{astrometry}
\end{minipage}
\end{center}
\end{figure}

We plot the constraints on companion mass as a function of semi-major axis afforded by the astrometric non-detections in Figure~\ref{astrometry}. Between $\approx1-2$~AU, the low RUWE allows us to exclude planets more massive than $\gtrsim$3~$M_J$. For wider separations, we use the method of \citet{Kervella2019} to calculate mass limits from the \textit{Hipparcos-Gaia} astrometry, from which we find that planets with masses as low as 1~$M_J$ can be excluded between $2.5-10$~AU. These constraints agree well with and marginally improve upon the non-detection of planets at projected separations above $>$2~AU in the MIRI imaging. However, we cannot independently confirm or reject close-separation planets with the existing astrometric data. Our existing astrometric constraints have low sensitivity to planetary-mass companions within $a<1$~AU.

If the MIRI infrared excess is indeed caused by a giant planet within 2~AU of WD~0310-688, it may be possible to detect the reflex orbital motion in the epoch astrometry that will be released in \textit{Gaia}~DR4. \citep{Sanderson2022} predict that \textit{Gaia} astrometry will lead to the detection of $8\pm2$ giant planets around white dwarfs; however, they have assumed that essentially none of these planets will be found within $<$2~AU as a result of destruction during stellar evolution.

\subsubsection{Doppler Monitoring}

In December 2023, we observed WD 0310-688 using the Magellan Carnegie Planet Finder Spectrograph \citep[PFS; ][]{2006SPIE.6269E..31C,2008SPIE.7014E..79C,2010SPIE.7735E..53C} to attempt Doppler monitoring for the planet candidate. Although WD 0310-688 is metal-polluted, only the Hydrogen Balmer lines were detectable in the PFS spectrum due to the low level of pollution. Using these lines, we achieved a precision of about 400\,m/s over 15\,minutes, or 200\,m/s per night with 1-hour monitoring. If orbiting at the minimum separation of 0.1\,AU, the planet must have a mass near or below the lower end of the self-luminous mass range (e.g., $<$1.1 M$_{\rm Jup}$, as determined in section \ref{sec:mass}), or it would be hotter than 248\,K. At this separation and mass, we would expect an RV semi-amplitude of 120\,m/s. Based on this initial observation, we conclude that detecting the planet via RV monitoring would require an intensive observation program, even if the planet is located in the most favorable part of the allowed parameter space. For the majority of the parameter space, this method is not feasible. Therefore, Doppler monitoring is not a practical method for confirming this planet.
However, RV confirmation may be feasible for white dwarf planet candidates with more metal lines in the visible spectrum or with shorter orbital periods \citep{2024MNRAS.527..977R}.

\subsubsection{Other Mid-IR Photometry}

We searched for archival data that could aid in illuminating the source of the infrared excess. We find that archival Spitzer data show no detectable IR excess. The absence of a Spitzer 8$\mu m$ (IRAC band 4) detection of infrared excess \citep{2007ApJS..171..206M} is consistent with our non detection of IR excess in the MIRI 7.7$\mu m$ band. In the WISE band 3 (12\,$\mu$m) we measure an IR excess of 16$\pm$14\%. Although this excess is consistent with our MIRI detection, it is also consistent with no excess at 1.1$\sigma$, so it provides little additional constraint on the system. In the WISE band 4 observations, the archival measurements do not provide sufficient sensitivity to produce any meaningful constraint on this source.

\section{Discussion} \label{sec:discuss}
We have detected infrared excess emission towards WD~0310-688, and we have shown that the emission is almost certainly coming from the white dwarf system. We can identify two likely explanations: a giant planet, or a cold debris disk. Here, we consider the two hypotheses, give arguments in each of their favor, and discuss the implications of each being correct. 

\subsection{Giant Planet} 

The first option we consider is a giant planet. It would have a radius similar to that of Jupiter, a mass of $3.0^{+5.5}_{-1.9}$\,M$_{\rm Jup}$, and an orbital separation between 0.1 and 2\,AU. The arguments in favor of this hypothesis are that
\begin{itemize}
    \item Planets of this mass are expected to exist around white dwarfs if they survive the star's red giant phase, particularly since A-type stars have higher rates of giant planet occurrence \citep{Reffert2015}.
    \item {A wide range of objects, from gas giant planets as small as Saturn to late M-dwarfs, have emitting areas similar to Jupiter, whereas debris disks can vary greatly in size. The radius inferred from a blackbody fit corresponds to the emitting area of a gas giant, which is suggestive. }
\end{itemize}
The arguments against this hypothesis are that
\begin{itemize}
    \item The emitting source is close to the white dwarf, which is expected to be rare. We know giant planets orbiting main sequence stars that are close enough to transit are rare \citep{2018MNRAS.474.4603V,2024arXiv240721743R}. Given our small sample size, one detection does not necessarily indicate a common phenomenon, but finding something so unusual so quickly is surprising. Nevertheless, the short-period giant planet WD~1856\,b exists \citep{Vanderburg_2020}, and this planet candidate could be a younger analog of that system's planet. Exoplanet astronomy has a history of discovering rare objects sooner than their actual occurrence rates would suggest \citep{2006Sci...314.1908G,2009Natur.462..891C,2015Natur.526..546V,2016Natur.533..221G}. 
    \item  Modeling the star's evolution (see Appendix \ref{AppA}) suggests that the planet is orbiting in a forbidden location \citep{Nordhaus2013MNRAS}, making it difficult to explain its existence. One explanation that has previously been invoked to explain short and intermediate period orbits around white dwarfs is common envelope evolution, in particular, scenarios involving additional energy sources beyond the planet's gravitational energy  \citep{2021MNRAS.501..676L,2021MNRAS.502L.110C, Merlov2021ApJL} or scenarios where the planet is engulfed at the end of the AGB phase, and mass loss facilitates the ejection of the envelope \citep{2024MNRAS.52711719Y}. However, we argue that a common envelope scenario is unlikely to be a good explanation for an intermediate-separation planet around WD 0310-688. It is already energetically difficult to make common envelope evolution work for WD~1856\,b. For this planet, which is at a much wider orbital separation then WD~1856\,b, this explanation would be even more challenging because amount of orbital energy available to eject the envelope is significantly less. Even greater amounts of additional energy would be required to prevent the planet from crashing into the core \citep{2021MNRAS.501..676L,2021MNRAS.502L.110C, Merlov2021ApJL}. 
    
    An alternative explanation for the presence of an intermediate period planet around WD 0310-688 is that the planet is currently undergoing tidal migration. However, in order for tidal migration to take place, the planet's eccentricity must be excited to very high values by additional massive objects in the system. There are no binary companions to this star that could drive tidal migration, however we observe numerous planet candidates at or beyond 80\,AU  distant in the MEOW images of WD 0310-688. Despite a high contamination rate from background AGN in our survey, if one of these sources is indeed a planet bound to the white dwarf, it could plausibly serve as a Kozai perturber, where the distant companion induces oscillations in the inclination and eccentricity of a planet's orbit, potentially leading to high eccentricity and subsequent tidal migration \citep{2020ApJ...904L...3M,2021ApJ...922....4S}.
    Planet-planet scattering is another plausible mechanism that could be responsible for the planet's current position \citep{2012ApJ...747..148D,2016MNRAS.458.3942V,2021MNRAS.501L..43M,2022MNRAS.512..104M}. Although close-in companions ($<80$\,AU) are absent, it is also plausible that an object may have been ejected from this system after interacting with our planet candidate. 
\end{itemize}

\subsection{Debris Disk}
Another possibility we consider is a disk. Converting an IR excess into precise debris disk parameters is challenging, but we can approach this in a couple of ways. First, we can assume the disk is optically thin and/or puffed up, similar to exozodiacal light. Under this assumption, and using the temperature of 248\,K from our blackbody fit for the dust, the semimajor axis would be 0.11\,AU, assuming an albedo of $\alpha = 0.06$, similar to that of interplanetary dust in our solar system \citep{2015ApJ...813...87Y}. The effective emitting area of the disk would be approximately the same as the surface area of Jupiter.

\begin{figure}[b]
\centering
\includegraphics[width=0.48\textwidth]{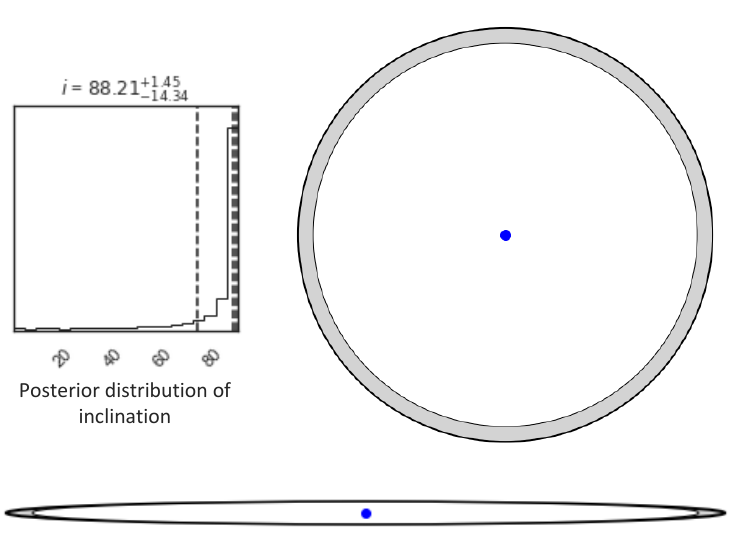}
\caption{Posterior of the disk inclination from the Jura disk model fit (upper left) and a schematic of the disk viewed at 88.21$^\circ$ (bottom) and face-on (upper right) for comparison.}
\label{DiskModel}
\end{figure}

Alternatively, we can assume an optically thick, thin, flat disk \citep{2003ApJ...584L..91J}. Using MCMC to fit the IR excess with the Jura disk model, we obtained the following parameters. In this case, the disk is a very skinny ring with inner radius of 171\,$R_{\rm WD}$, outer radius 185\,$R_{\rm WD}$, and is inclined nearly edge-on (see Figure \ref{DiskModel}). 

The arguments in favor of this hypothesis are that: 
\begin{itemize}
    \item Dust disks are common around white dwarfs, and although we have not previously observed one at such a cold temperature \citep{2010ApJ...714.1386F}, this is likely due to the lack of sensitivity before the advent of JWST. It is often more likely to encounter an unusual manifestation of a common phenomenon, like a debris disk, than a typical presentation of an uncommon phenomenon, such as a planet in an unlikely orbital configuration forbidden zone. This aligns with the principle of considering the most probable explanations before exploring less likely alternatives.
    \item  The white dwarf is polluted as discussed in section \ref{pollution}, and a debris disk is the most probable source of the pollution (although we note that pollution does not preclude the presence of a close-in planets \citealt{2019Natur.576...61G}). {Infrared observations indicate white dwarfs with observed} debris disks are heavily polluted, typically with total accretion rate $>3\times10^8$\,g/s \citep{2009ApJ...694..805F}, much higher than the accretion rate we measure for WD~0310-688.
\end{itemize}
The arguments against this hypothesis
\begin{itemize}
    \item In order to explain the IR emission with a disk model, we must finely tune the disk's geometry to give a surface area similar to Jupiter. In particular, best-fit parameters for an optically thick disk seem a bit contrived, with an extremely thin ring (only 7\% of the disk's radial extent) and a nearly edge-on inclination. However, other disks have been found with similarly thin-ring geometries \citep{2022ApJ...939..108B}, and there is no reason that such disks could not be precisely aligned with our line of sight, so such a seemingly unusual geometry cannot be ruled out.   
    \item  The location of the dust in the optically thin case is well outside the Roche Limit, which would be contrary to our normal picture of white dwarf disks and pollution (although see the discovery of transiting systems outside the Roche radius for the rocky material that typically makes up dusty debris disks \citealt{2020ApJ...897..171V,2022MNRAS.511.1647F}). 
\end{itemize}

\subsection{Implications of the Discovery} 

Regardless of the underlying cause, this IR excess is expanding our understanding of white dwarf planetary systems.

\subsubsection{If it is a Planet} 
Confirming this source as an exoplanet would be particularly exciting for several reasons. It represents the closest planet candidate around a white dwarf discovered to date, at a distance of 10.4 parsecs. It is the first planet candidate identified using the infrared excess detection technique. It is the first white dwarf planet candidate discovered at an intermediate separation (0.1-2\,AU). If confirmed, it would be one of the coldest worlds for which direct spectral atmospheric characterization is possible.

Furthermore, the host is a remnant of an A or late B star, and the planet would offer critical demographic constraints on this population, which is challenging to study by other means. With few exceptions, planets around A- and B-type main sequence stars are generally accessible only through high-contrast imaging. Searches for such planets have yielded several detections \citep[e.g.][]{Lagrange2010,Janson2021bcen}, but are typically limited to planets at relatively large semi-major axes \citep[tens or hundreds of au;][]{Nielsen2019,Vigan2021,2021A&A...646A.164J}. Past the main sequence, in the red giant phase, the stars have expanded and cooled sufficiently to be better suitable for RV studies. Such studies have also yielded several planet candidates in the relevant stellar mass range \citep[e.g.][]{Johnson2007,Reffert2015}, though in this case the detection range is limited to small separations in the range of typically a few au, and the actual estimation of stellar mass is subject to large uncertainties \citep[e.g.][]{Lloyd2011,Johnson2013}. As we have seen above, white dwarf imaging with JWST can cover planets over the whole separation space from the Roche limit (via infrared excess) up to hundreds of au (via resolved imaging) simultaneously, providing a much more complete demographic overview for the stellar remnant phase of massive stars. Comparisons with the main sequence and red giant demographics can also provide the first constraints on how planetary systems are affected through the late stages of stellar evolution.

\subsubsection{If it is a disk:} 
Only three WD disks have previous Spitzer/MIPS measurements: G29-38 \citep{2005ApJ...635L.161R}, GD 56 and GD 133 \citep{2007ApJ...663.1285J}. Few distant and cold disk candidates have been identified around hot, bright white dwarfs with Spitzer at 24\,$\mu m$. A notable instance is the central star of the Helix Nebula \citep{2007ApJ...657L..41S}. However, their interpretation as cold dust disks is complicated by dusty outflows associated with their immediate progenitors and binarity \citep{2011AJ....142...75C,2014AJ....147..142C}. If this source is actually a disk rather than a planet, it would be {perhaps unusual} compared with other known white dwarf disks and could significantly expand our understanding of how planetary debris is accreted onto polluted white dwarf stars \citep{2020NatAs...4..328C}.

If confirmed as a debris disk, this system, alongside PG 1225-079 \citep{2010ApJ...714.1386F}, would represent one of the coldest debris disks ever detected around a white dwarf. These disks may signify distinct evolutionary stages--either at the nascent phase, where the disk is just beginning to form, or nearing the terminal phase, where the disk is predominantly dissipated. \citet{2017MNRAS.468..154B} predicted that the JWST would uncover numerous small debris disks, suggesting that such findings may be more common than previously anticipated.

Finally, considering our estimated accretion rate, if this source is indeed a debris disk, it would represent the white dwarf system with the lowest known accretion rate associated with a detectable disk.

\subsection{How to distinguish the two scenarios} 
The existing data on this system does not provide enough information to definitely differentiate between the planet and disk hypotheses. We surmise that the easiest way to differentiate between the two will be to obtain a MIRI MRS spectrum of the blended SED. Even in the case of a cloudy planet, several hours of MRS time enables the detection of atmospheric features in an exoplanet's atmosphere that would provide concrete evidence of a planet. Conversely, a lack of detection of planetary atmospheric features consistent with an exoplanet atmosphere would provide strong evidence that the source of infrared excess is instead a cold dust disk. If it is indeed a debris disk, very likely it will have 10 micron silicate feature, which has been seen in white dwarfs disks with mid-infrared data \citep{2009AJ....137.3191J,2024MNRAS.529L..41S}.

\section{Conclusions} \label{sec:conclude}
 
In this paper, we described the detection of infrared excess around the white dwarf WD 0310-688 and found that:
\begin{itemize}
\item The IR excess is best fit by a blackbody with a temperature of 248$^{+84}_{-61}$\,K and an emitting area of 1.03$^{+0.67}_{-0.35}$\,R$_{\rm Jup}$.
\item The source of the IR excess is constrained to be within 0.1-2\,AU, corresponding to orbital periods of 14 days to 3.4 years.
\item If the emitting source is an exoplanet, its mass is constrained to $3.0^{+5.5}_{-1.9}$\,M$_{\rm Jup}$, though it could be as low as 0.2\,M$_{\rm Jup}$ if the planet's heat comes from irradiation.
\end{itemize}

The detection of a source, whether it is a planet or a cold dust disk, at this separation is unprecedented. It is one of only a handful of known white dwarf planets or planet candidates. It represents the nearest planet candidate around a white dwarf discovered to date, at a distance of 10.4 parsecs. It is the first planet candidate ever identified using the infrared excess detection technique. This technique is well suited for the detection of white dwarf exoplanets, and may reveal many more candidates during JWST's lifetime.

It is the first white dwarf planet candidate discovered at an intermediate separation (0.1-2 AU). This is particularly significant because planets at this separation are expected to have been destroyed during the star’s red giant phase. Therefore, if confirmed, the planet must have migrated to its current location after the star evolved into a white dwarf. Confirming and further study of this planet will be crucial for understanding post-main sequence planetary evolution and the fate of planets as their host stars die.

We determine that follow-up observations with JWST MRS will enable us to confirm the source of the IR excess as a planet and rule out the disk hypotheses. If confirmed, it will be one of the coldest worlds for which direct spectral atmospheric characterization is possible. Subsequent spectroscopic observations could allow for comparative studies with the coldest directly imaged free-floating planetary-mass worlds (such as the mid Y-dwarf WISE 0855-0714 \citep{2024AJ....167....5L}, which is comparable in mass, age, and temperature).

\section*{Acknowledgement}
We would like to thank Misty Cracraft, Karl Gordon and the JWST Helpdesk for support with MIRI calibration files and absolute flux calibrations. This research has made use of the SIMBAD database,
operated at CDS, Strasbourg, France \citep{2000A&AS..143....9W}.
This research has made use of the NASA Exoplanet Archive, which is operated by the California Institute of Technology, under contract with the National Aeronautics and Space Administration under the Exoplanet Exploration Program. This paper leverages data gathered with the 6.5 meter Magellan Telescopes located at Las Campanas Observatory, Chile.
This work is based [in part] on observations made with the NASA/ESA/CSA James Webb Space Telescope. The data were obtained from the Mikulski Archive for Space Telescopes at the Space Telescope Science Institute, which is operated by the Association of Universities for Research in Astronomy, Inc., under NASA contract NAS 5-03127 for JWST. These observations are associated with program \#4403.

\facilities{\it JWST, HST, Gaia, Hipparcos, Magellan, Spitzer, WISE}. All the JWST data used in this paper can be found in MAST:  \dataset[10.17909/cjkx-kp07]{http://dx.doi.org/10.17909/cjkx-kp07}.

\software{{\tt astro.py} \citep{2013A&A...558A..33A, 2018AJ....156..123A, 2022ApJ...935..167A}, {\tt numpy.py} \citep{5725236}, {\tt wdwarfdate} \citep{2022AJ....164...62K}, {\tt POSEIDON} \citep{2017MNRAS.469.1979M,2023JOSS....8.4873M}, \texttt{MESA} version v7503; \citealt{Dotter2016}, \citealt{Choi2016}, \citep{Paxton2011,Paxton2013, Paxton2015}, {\tt SAOImage DS9} \citep{2000ascl.soft03002S}, \texttt{VIP} \citep{VIP}, {\tt corner.py} \citep{corner} and {\tt ChatGPT} was utilized to improve wording at the sentence level and assist with coding inquires; Last accessed in August 2024.}

\appendix
\section{The Implications of Stellar Evolution on a Planet's Orbit}\label{AppA}

\begin{figure*}[h]
\centering
\includegraphics[width=0.49\textwidth]{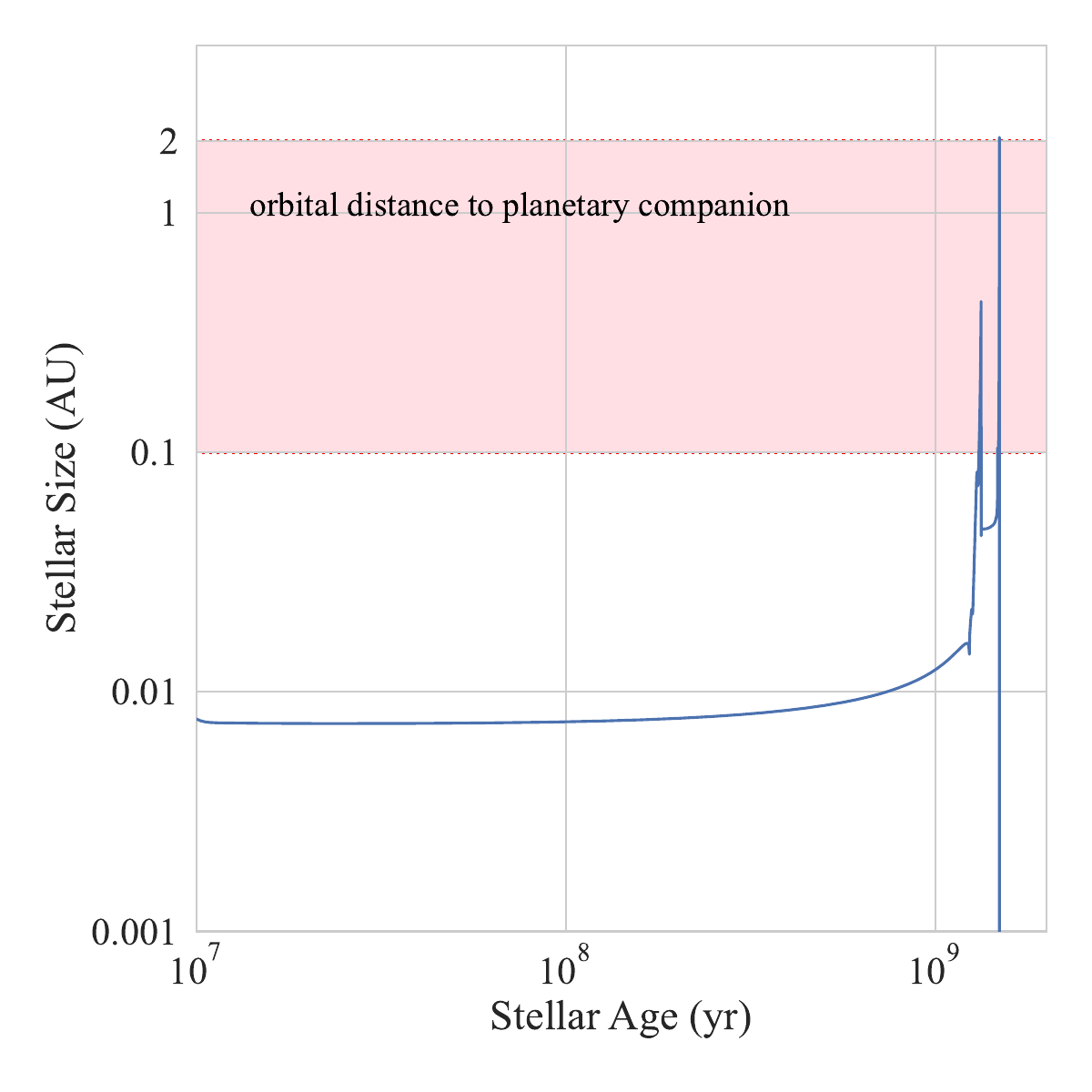}
\caption{Evolution of the stellar radius as a function of age for a 1.9\,$M_{\odot}$ star, simulated using \texttt{MESA} Isochrones \& Stellar Tracks (\texttt{MESA} version v7503). The shaded pink region represents the range of possible orbital separations of the planetary companion (0.1-2\,AU). The dotted lines mark the bounds of this range. The maximum radius reached by the star at the tip of the red giant branch (2.1\,AU) exceeds even the most generous estimate of the orbital separation, indicating that the planetary companion likely falls into the forbidden zone.}
\label{fig:size}
\end{figure*}

We can draw insights as to whether the planetary companion falls into the forbidden zone by investigating the current orbital separation in comparison to the size of the host when it was at the tip of the red giant branch (RGB). To this end, we simulated the evolution of a 1.9\,$M_{\odot}$ star using \texttt{MESA} Isochrones \& Stellar Tracks\footnote{\url{https://waps.cfa.harvard.edu/MIST/}} (\texttt{MESA} version v7503; \citealt{Dotter2016}, \citealt{Choi2016}, \citealt{Paxton2011,Paxton2013, Paxton2015}). We investigated a range of initial metallicities, determining that the orbital separation of the planet was within the maximum RGB radius in all cases. In Figure~\ref{fig:size}, we illustrate the change in stellar size as a function of age. The estimated bounds of the orbital separation of the companion, 0.1-2\,AU, are shown as dotted lines and the intermediary range between these extremes is highlighted in pink. Note that even the most generous orbital separation estimate is less than the maximum RGB size of the host (2.1\,AU).

\bibliography{main}{}

\begin{thebibliography}{}
\expandafter\ifx\csname natexlab\endcsname\relax\def\natexlab#1{#1}\fi
\providecommand{\url}[1]{\href{#1}{#1}}
\providecommand{\dodoi}[1]{doi:~\href{http://doi.org/#1}{\nolinkurl{#1}}}
\providecommand{\doeprint}[1]{\href{http://ascl.net/#1}{\nolinkurl{http://ascl.net/#1}}}
\providecommand{\doarXiv}[1]{\href{https://arxiv.org/abs/#1}{\nolinkurl{https://arxiv.org/abs/#1}}}

\bibitem[{{Agol}(2011)}]{2011ApJ...731L..31A}
{Agol}, E. 2011, ApJL, 731, L31, \dodoi{10.1088/2041-8205/731/2/L31}

\bibitem[{{Alcock} {et~al.}(1986){Alcock}, {Fristrom}, \& {Siegelman}}]{1986ApJ...302..462A}
{Alcock}, C., {Fristrom}, C.~C., \& {Siegelman}, R. 1986, \apj, 302, 462, \dodoi{10.1086/164005}

\bibitem[{{All{\`e}gre} {et~al.}(1995){All{\`e}gre}, {Poirier}, {Humler}, \& {Hofmann}}]{1995E&PSL.134..515A}
{All{\`e}gre}, C.~J., {Poirier}, J.-P., {Humler}, E., \& {Hofmann}, A.~W. 1995, Earth and Planetary Science Letters, 134, 515, \dodoi{10.1016/0012-821X(95)00123-T}

\bibitem[{{Astropy Collaboration} {et~al.}(2013){Astropy Collaboration}, {Robitaille}, {Tollerud}, {Greenfield}, {Droettboom}, {Bray}, {Aldcroft}, {Davis}, {Ginsburg}, {Price-Whelan}, {Kerzendorf}, {Conley}, {Crighton}, {Barbary}, {Muna}, {Ferguson}, {Grollier}, {Parikh}, {Nair}, {Unther}, {Deil}, {Woillez}, {Conseil}, {Kramer}, {Turner}, {Singer}, {Fox}, {Weaver}, {Zabalza}, {Edwards}, {Azalee Bostroem}, {Burke}, {Casey}, {Crawford}, {Dencheva}, {Ely}, {Jenness}, {Labrie}, {Lim}, {Pierfederici}, {Pontzen}, {Ptak}, {Refsdal}, {Servillat}, \& {Streicher}}]{2013A&A...558A..33A}
{Astropy Collaboration}, {Robitaille}, T.~P., {Tollerud}, E.~J., {et~al.} 2013, \aap, 558, A33, \dodoi{10.1051/0004-6361/201322068}

\bibitem[{{Astropy Collaboration} {et~al.}(2018){Astropy Collaboration}, {Price-Whelan}, {Sip{\H{o}}cz}, {G{\"u}nther}, {Lim}, {Crawford}, {Conseil}, {Shupe}, {Craig}, {Dencheva}, {Ginsburg}, {VanderPlas}, {Bradley}, {P{\'e}rez-Su{\'a}rez}, {de Val-Borro}, {Aldcroft}, {Cruz}, {Robitaille}, {Tollerud}, {Ardelean}, {Babej}, {Bach}, {Bachetti}, {Bakanov}, {Bamford}, {Barentsen}, {Barmby}, {Baumbach}, {Berry}, {Biscani}, {Boquien}, {Bostroem}, {Bouma}, {Brammer}, {Bray}, {Breytenbach}, {Buddelmeijer}, {Burke}, {Calderone}, {Cano Rodr{\'\i}guez}, {Cara}, {Cardoso}, {Cheedella}, {Copin}, {Corrales}, {Crichton}, {D'Avella}, {Deil}, {Depagne}, {Dietrich}, {Donath}, {Droettboom}, {Earl}, {Erben}, {Fabbro}, {Ferreira}, {Finethy}, {Fox}, {Garrison}, {Gibbons}, {Goldstein}, {Gommers}, {Greco}, {Greenfield}, {Groener}, {Grollier}, {Hagen}, {Hirst}, {Homeier}, {Horton}, {Hosseinzadeh}, {Hu}, {Hunkeler}, {Ivezi{\'c}}, {Jain}, {Jenness}, {Kanarek}, {Kendrew}, {Kern}, {Kerzendorf}, {Khvalko}, {King}, {Kirkby}, {Kulkarni},
  {Kumar}, {Lee}, {Lenz}, {Littlefair}, {Ma}, {Macleod}, {Mastropietro}, {McCully}, {Montagnac}, {Morris}, {Mueller}, {Mumford}, {Muna}, {Murphy}, {Nelson}, {Nguyen}, {Ninan}, {N{\"o}the}, {Ogaz}, {Oh}, {Parejko}, {Parley}, {Pascual}, {Patil}, {Patil}, {Plunkett}, {Prochaska}, {Rastogi}, {Reddy Janga}, {Sabater}, {Sakurikar}, {Seifert}, {Sherbert}, {Sherwood-Taylor}, {Shih}, {Sick}, {Silbiger}, {Singanamalla}, {Singer}, {Sladen}, {Sooley}, {Sornarajah}, {Streicher}, {Teuben}, {Thomas}, {Tremblay}, {Turner}, {Terr{\'o}n}, {van Kerkwijk}, {de la Vega}, {Watkins}, {Weaver}, {Whitmore}, {Woillez}, {Zabalza}, \& {Astropy Contributors}}]{2018AJ....156..123A}
{Astropy Collaboration}, {Price-Whelan}, A.~M., {Sip{\H{o}}cz}, B.~M., {et~al.} 2018, \aj, 156, 123, \dodoi{10.3847/1538-3881/aabc4f}

\bibitem[{{Astropy Collaboration} {et~al.}(2022){Astropy Collaboration}, {Price-Whelan}, {Lim}, {Earl}, {Starkman}, {Bradley}, {Shupe}, {Patil}, {Corrales}, {Brasseur}, {N{\"o}the}, {Donath}, {Tollerud}, {Morris}, {Ginsburg}, {Vaher}, {Weaver}, {Tocknell}, {Jamieson}, {van Kerkwijk}, {Robitaille}, {Merry}, {Bachetti}, {G{\"u}nther}, {Aldcroft}, {Alvarado-Montes}, {Archibald}, {B{\'o}di}, {Bapat}, {Barentsen}, {Baz{\'a}n}, {Biswas}, {Boquien}, {Burke}, {Cara}, {Cara}, {Conroy}, {Conseil}, {Craig}, {Cross}, {Cruz}, {D'Eugenio}, {Dencheva}, {Devillepoix}, {Dietrich}, {Eigenbrot}, {Erben}, {Ferreira}, {Foreman-Mackey}, {Fox}, {Freij}, {Garg}, {Geda}, {Glattly}, {Gondhalekar}, {Gordon}, {Grant}, {Greenfield}, {Groener}, {Guest}, {Gurovich}, {Handberg}, {Hart}, {Hatfield-Dodds}, {Homeier}, {Hosseinzadeh}, {Jenness}, {Jones}, {Joseph}, {Kalmbach}, {Karamehmetoglu}, {Ka{\l}uszy{\'n}ski}, {Kelley}, {Kern}, {Kerzendorf}, {Koch}, {Kulumani}, {Lee}, {Ly}, {Ma}, {MacBride}, {Maljaars}, {Muna}, {Murphy}, {Norman},
  {O'Steen}, {Oman}, {Pacifici}, {Pascual}, {Pascual-Granado}, {Patil}, {Perren}, {Pickering}, {Rastogi}, {Roulston}, {Ryan}, {Rykoff}, {Sabater}, {Sakurikar}, {Salgado}, {Sanghi}, {Saunders}, {Savchenko}, {Schwardt}, {Seifert-Eckert}, {Shih}, {Jain}, {Shukla}, {Sick}, {Simpson}, {Singanamalla}, {Singer}, {Singhal}, {Sinha}, {Sip{\H{o}}cz}, {Spitler}, {Stansby}, {Streicher}, {{\v{S}}umak}, {Swinbank}, {Taranu}, {Tewary}, {Tremblay}, {de Val-Borro}, {Van Kooten}, {Vasovi{\'c}}, {Verma}, {de Miranda Cardoso}, {Williams}, {Wilson}, {Winkel}, {Wood-Vasey}, {Xue}, {Yoachim}, {Zhang}, {Zonca}, \& {Astropy Project Contributors}}]{2022ApJ...935..167A}
{Astropy Collaboration}, {Price-Whelan}, A.~M., {Lim}, P.~L., {et~al.} 2022, \apj, 935, 167, \dodoi{10.3847/1538-4357/ac7c74}

\bibitem[{{Ballering} {et~al.}(2022){Ballering}, {Levens}, {Su}, \& {Cleeves}}]{2022ApJ...939..108B}
{Ballering}, N.~P., {Levens}, C.~I., {Su}, K. Y.~L., \& {Cleeves}, L.~I. 2022, \apj, 939, 108, \dodoi{10.3847/1538-4357/ac9a4a}

\bibitem[{{Barber} {et~al.}(2012){Barber}, {Patterson}, {Kilic}, {Leggett}, {Dufour}, {Bloom}, \& {Starr}}]{2012ApJ...760...26B}
{Barber}, S.~D., {Patterson}, A.~J., {Kilic}, M., {et~al.} 2012, ApJ, 760, 26, \dodoi{10.1088/0004-637X/760/1/26}

\bibitem[{{Barnes} \& {Heller}(2013)}]{2013AsBio..13..279B}
{Barnes}, R., \& {Heller}, R. 2013, Astrobiology, 13, 279, \dodoi{10.1089/ast.2012.0867}

\bibitem[{{Becker} {et~al.}(2023){Becker}, {Seligman}, {Adams}, \& {Styczinski}}]{2023ApJ...945L..24B}
{Becker}, J., {Seligman}, D.~Z., {Adams}, F.~C., \& {Styczinski}, M.~J. 2023, \apjl, 945, L24, \dodoi{10.3847/2041-8213/acbe44}

\bibitem[{{Becklin} {et~al.}(2005){Becklin}, {Farihi}, {Jura}, {Song}, {Weinberger}, \& {Zuckerman}}]{2005ApJ...632L.119B}
{Becklin}, E.~E., {Farihi}, J., {Jura}, M., {et~al.} 2005, ApJL, 632, L119, \dodoi{10.1086/497826}

\bibitem[{{B{\'e}dard} {et~al.}(2017){B{\'e}dard}, {Bergeron}, \& {Fontaine}}]{2017ApJ...848...11B}
{B{\'e}dard}, A., {Bergeron}, P., \& {Fontaine}, G. 2017, \apj, 848, 11, \dodoi{10.3847/1538-4357/aa8bb6}

\bibitem[{{Belokurov} {et~al.}(2020){Belokurov}, {Penoyre}, {Oh}, {Iorio}, {Hodgkin}, {Evans}, {Everall}, {Koposov}, {Tout}, {Izzard}, {Clarke}, \& {Brown}}]{Belokurov20}
{Belokurov}, V., {Penoyre}, Z., {Oh}, S., {et~al.} 2020, \mnras, 496, 1922, \dodoi{10.1093/mnras/staa1522}

\bibitem[{{Blackman} {et~al.}(2024){Blackman}, {Danielski}, {Bachelet}, {Beaulieu}, {Beichman}, {Bennett}, {Bhattacharya}, {Cole}, {Koshimoto}, {Ranc}, {Rektsini}, {Terry}, \& {Vandorou}}]{2024jwst.prop.6078B}
{Blackman}, J., {Danielski}, C., {Bachelet}, E., {et~al.} 2024, {Confirmation of a Jovian Planet Analog Orbiting a White Dwarf, Rare Low-mass Neutron Star or Black Hole}, JWST Proposal. Cycle 3, ID. \#6078

\bibitem[{{Blackman} {et~al.}(2021){Blackman}, {Beaulieu}, {Bennett}, {Danielski}, {Alard}, {Cole}, {Vandorou}, {Ranc}, {Terry}, {Bhattacharya}, {Bond}, {Bachelet}, {Veras}, {Koshimoto}, {Batista}, \& {Marquette}}]{2021Natur.598..272B}
{Blackman}, J.~W., {Beaulieu}, J.~P., {Bennett}, D.~P., {et~al.} 2021, Nature, 598, 272, \dodoi{10.1038/s41586-021-03869-6}

\bibitem[{{Blouin} {et~al.}(2018{\natexlab{a}}){Blouin}, {Dufour}, \& {Allard}}]{blouin2018a}
{Blouin}, S., {Dufour}, P., \& {Allard}, N.~F. 2018{\natexlab{a}}, ApJ, 863, 184, \dodoi{10.3847/1538-4357/aad4a9}

\bibitem[{{Blouin} {et~al.}(2018{\natexlab{b}}){Blouin}, {Dufour}, {Allard}, \& {Kilic}}]{blouin2018b}
{Blouin}, S., {Dufour}, P., {Allard}, N.~F., \& {Kilic}, M. 2018{\natexlab{b}}, ApJ, 867, 161, \dodoi{10.3847/1538-4357/aae53a}

\bibitem[{{Blouin} \& {Xu}(2022)}]{2022MNRAS.510.1059B}
{Blouin}, S., \& {Xu}, S. 2022, \mnras, 510, 1059, \dodoi{10.1093/mnras/stab3446}

\bibitem[{{Bonsor} {et~al.}(2017){Bonsor}, {Farihi}, {Wyatt}, \& {van Lieshout}}]{2017MNRAS.468..154B}
{Bonsor}, A., {Farihi}, J., {Wyatt}, M.~C., \& {van Lieshout}, R. 2017, \mnras, 468, 154, \dodoi{10.1093/mnras/stx425}

\bibitem[{{Bouchet} {et~al.}(2015){Bouchet}, {Garc{\'\i}a-Mar{\'\i}n}, {Lagage}, {Amiaux}, {Augu{\'e}res}, {Bauwens}, {Blommaert}, {Chen}, {Detre}, {Dicken}, {Dubreuil}, {Galdemard}, {Gastaud}, {Glasse}, {Gordon}, {Gougnaud}, {Guillard}, {Justtanont}, {Krause}, {Leboeuf}, {Longval}, {Martin}, {Mazy}, {Moreau}, {Olofsson}, {Ray}, {Rees}, {Renotte}, {Ressler}, {Ronayette}, {Salasca}, {Scheithauer}, {Sykes}, {Thelen}, {Wells}, {Wright}, \& {Wright}}]{2015PASP..127..612B}
{Bouchet}, P., {Garc{\'\i}a-Mar{\'\i}n}, M., {Lagage}, P.~O., {et~al.} 2015, \pasp, 127, 612, \dodoi{10.1086/682254}

\bibitem[{{Brandt}(2018)}]{Brandt2018}
{Brandt}, T.~D. 2018, \apjs, 239, 31, \dodoi{10.3847/1538-4365/aaec06}

\bibitem[{{Brandt}(2021)}]{Brandt2021}
---. 2021, \apjs, 254, 42, \dodoi{10.3847/1538-4365/abf93c}

\bibitem[{{Chamandy} {et~al.}(2021){Chamandy}, {Blackman}, {Nordhaus}, \& {Wilson}}]{2021MNRAS.502L.110C}
{Chamandy}, L., {Blackman}, E.~G., {Nordhaus}, J., \& {Wilson}, E. 2021, MNRAS, 502, L110, \dodoi{10.1093/mnrasl/slab017}

\bibitem[{{Charbonneau} {et~al.}(2009){Charbonneau}, {Berta}, {Irwin}, {Burke}, {Nutzman}, {Buchhave}, {Lovis}, {Bonfils}, {Latham}, {Udry}, {Murray-Clay}, {Holman}, {Falco}, {Winn}, {Queloz}, {Pepe}, {Mayor}, {Delfosse}, \& {Forveille}}]{2009Natur.462..891C}
{Charbonneau}, D., {Berta}, Z.~K., {Irwin}, J., {et~al.} 2009, \nat, 462, 891, \dodoi{10.1038/nature08679}

\bibitem[{{Chen} {et~al.}(2020){Chen}, {Su}, \& {Xu}}]{2020NatAs...4..328C}
{Chen}, C.~H., {Su}, K. Y.~L., \& {Xu}, S. 2020, Nature Astronomy, 4, 328, \dodoi{10.1038/s41550-020-1067-6}

\bibitem[{{Cheng} {et~al.}(2024){Cheng}, {Schlaufman}, \& {Caiazzo}}]{2024jwst.prop.6410C}
{Cheng}, S., {Schlaufman}, K., \& {Caiazzo}, I. 2024, {A Giant Planet Candidate Orbiting a Young, Massive White Dwarf}, JWST Proposal. Cycle 3, ID. \#6410

\bibitem[{{Choi} {et~al.}(2016){Choi}, {Dotter}, {Conroy}, {Cantiello}, {Paxton}, \& {Johnson}}]{Choi2016}
{Choi}, J., {Dotter}, A., {Conroy}, C., {et~al.} 2016, \apj, 823, 102, \dodoi{10.3847/0004-637X/823/2/102}

\bibitem[{{Chu} {et~al.}(2011){Chu}, {Su}, {Bilikova}, {Gruendl}, {De Marco}, {Guerrero}, {Updike}, {Volk}, \& {Rauch}}]{2011AJ....142...75C}
{Chu}, Y.-H., {Su}, K. Y.~L., {Bilikova}, J., {et~al.} 2011, \aj, 142, 75, \dodoi{10.1088/0004-6256/142/3/75}

\bibitem[{{Clayton} {et~al.}(2014){Clayton}, {De Marco}, {Nordhaus}, {Green}, {Rauch}, {Werner}, \& {Chu}}]{2014AJ....147..142C}
{Clayton}, G.~C., {De Marco}, O., {Nordhaus}, J., {et~al.} 2014, \aj, 147, 142, \dodoi{10.1088/0004-6256/147/6/142}

\bibitem[{{Crane} {et~al.}(2006){Crane}, {Shectman}, \& {Butler}}]{2006SPIE.6269E..31C}
{Crane}, J.~D., {Shectman}, S.~A., \& {Butler}, R.~P. 2006, in Society of Photo-Optical Instrumentation Engineers (SPIE) Conference Series, Vol. 6269, Ground-based and Airborne Instrumentation for Astronomy, ed. I.~S. {McLean} \& M.~{Iye}, 626931, \dodoi{10.1117/12.672339}

\bibitem[{{Crane} {et~al.}(2010){Crane}, {Shectman}, {Butler}, {Thompson}, {Birk}, {Jones}, \& {Burley}}]{2010SPIE.7735E..53C}
{Crane}, J.~D., {Shectman}, S.~A., {Butler}, R.~P., {et~al.} 2010, in Society of Photo-Optical Instrumentation Engineers (SPIE) Conference Series, Vol. 7735, Ground-based and Airborne Instrumentation for Astronomy III, ed. I.~S. {McLean}, S.~K. {Ramsay}, \& H.~{Takami}, 773553, \dodoi{10.1117/12.857792}

\bibitem[{{Crane} {et~al.}(2008){Crane}, {Shectman}, {Butler}, {Thompson}, \& {Burley}}]{2008SPIE.7014E..79C}
{Crane}, J.~D., {Shectman}, S.~A., {Butler}, R.~P., {Thompson}, I.~B., \& {Burley}, G.~S. 2008, in Society of Photo-Optical Instrumentation Engineers (SPIE) Conference Series, Vol. 7014, Ground-based and Airborne Instrumentation for Astronomy II, ed. I.~S. {McLean} \& M.~M. {Casali}, 701479, \dodoi{10.1117/12.789637}

\bibitem[{{Cunningham} {et~al.}(2022){Cunningham}, {Wheatley}, {Tremblay}, {G{\"a}nsicke}, {King}, {Toloza}, \& {Veras}}]{2022Natur.602..219C}
{Cunningham}, T., {Wheatley}, P.~J., {Tremblay}, P.-E., {et~al.} 2022, Nature, 602, 219, \dodoi{10.1038/s41586-021-04300-w}

\bibitem[{{Cutri} {et~al.}(2003){Cutri}, {Skrutskie}, {van Dyk}, {Beichman}, {Carpenter}, {Chester}, {Cambresy}, {Evans}, {Fowler}, {Gizis}, {Howard}, {Huchra}, {Jarrett}, {Kopan}, {Kirkpatrick}, {Light}, {Marsh}, {McCallon}, {Schneider}, {Stiening}, {Sykes}, {Weinberg}, {Wheaton}, {Wheelock}, \& {Zacarias}}]{2003yCat.2246....0C}
{Cutri}, R.~M., {Skrutskie}, M.~F., {van Dyk}, S., {et~al.} 2003, {VizieR Online Data Catalog: 2MASS All-Sky Catalog of Point Sources (Cutri+ 2003)}, VizieR On-line Data Catalog: II/246. Originally published in: 2003yCat.2246....0C

\bibitem[{{Debes} {et~al.}(2024){Debes}, {Sankrit}, {Fischer}, {Frazer}, {Hirschauer}, {Rowlands}, {Burger}, {Swaters}, {Jedrzejewski}, {Gomez}, {Dos Santos}, {Hernandez}, {Miller}, {Payne}, {Rafelski}, {Wevers}, {Anderson}, {Bair}, {Bello}, {Carlberg}, {Charlow}, {Cortese}, {Dencheva}, {Ellis}, {Falk}, {Fleming}, {Forshay}, {Gilani}, {Hall}, {Kimball}, {Kelley}, {Kidwell}, {Kotler}, {Kovacs}, {James}, {Rahmani}, {Rodriguez}, {Roman-Duval}, {Soderblom}, {Sherbert}, {Welty}, \& {Wolfe}}]{2024cos..rept....1D}
{Debes}, J., {Sankrit}, R., {Fischer}, T., {et~al.} 2024, {The Hubble Advanced Spectral Product (HASP) Program}, Instrument Science Report COS 2024-01, 31 pages

\bibitem[{{Debes} \& {Sigurdsson}(2002)}]{2002ApJ...572..556D}
{Debes}, J.~H., \& {Sigurdsson}, S. 2002, ApJ, 572, 556, \dodoi{10.1086/340291}

\bibitem[{{Debes} {et~al.}(2005){Debes}, {Sigurdsson}, \& {Woodgate}}]{2005ApJ...633.1168D}
{Debes}, J.~H., {Sigurdsson}, S., \& {Woodgate}, B.~E. 2005, \apj, 633, 1168, \dodoi{10.1086/491640}

\bibitem[{{Debes} {et~al.}(2012){Debes}, {Walsh}, \& {Stark}}]{2012ApJ...747..148D}
{Debes}, J.~H., {Walsh}, K.~J., \& {Stark}, C. 2012, \apj, 747, 148, \dodoi{10.1088/0004-637X/747/2/148}

\bibitem[{{Dotter}(2016)}]{Dotter2016}
{Dotter}, A. 2016, \apjs, 222, 8, \dodoi{10.3847/0067-0049/222/1/8}

\bibitem[{{Dufour} {et~al.}(2017){Dufour}, {Blouin}, {Coutu}, {Fortin-Archambault}, {Thibeault}, {Bergeron}, \& {Fontaine}}]{Dufour2017}
{Dufour}, P., {Blouin}, S., {Coutu}, S., {et~al.} 2017, in Astronomical Society of the Pacific Conference Series, Vol. 509, 20th European White Dwarf Workshop, ed. P.~E. {Tremblay}, B.~{Gaensicke}, \& T.~{Marsh}, 3, \dodoi{10.48550/arXiv.1610.00986}

\bibitem[{{Dufour} {et~al.}(2010){Dufour}, {Kilic}, {Fontaine}, {Bergeron}, {Lachapelle}, {Kleinman}, \& {Leggett}}]{2010ApJ...719..803D}
{Dufour}, P., {Kilic}, M., {Fontaine}, G., {et~al.} 2010, \apj, 719, 803, \dodoi{10.1088/0004-637X/719/1/803}

\bibitem[{{Endl} \& {Williams}(2018)}]{Endl2018}
{Endl}, M., \& {Williams}, K. 2018, in The 21st European Workshop on White Dwarfs, 1--2, \dodoi{hdl.handle.net/2152/71591}

\bibitem[{{Farihi} {et~al.}(2008{\natexlab{a}}){Farihi}, {Becklin}, \& {Zuckerman}}]{2008ApJ...681.1470F}
{Farihi}, J., {Becklin}, E.~E., \& {Zuckerman}, B. 2008{\natexlab{a}}, ApJ, 681, 1470, \dodoi{10.1086/588726}

\bibitem[{{Farihi} {et~al.}(2010){Farihi}, {Jura}, {Lee}, \& {Zuckerman}}]{2010ApJ...714.1386F}
{Farihi}, J., {Jura}, M., {Lee}, J.~E., \& {Zuckerman}, B. 2010, \apj, 714, 1386, \dodoi{10.1088/0004-637X/714/2/1386}

\bibitem[{{Farihi} {et~al.}(2009){Farihi}, {Jura}, \& {Zuckerman}}]{2009ApJ...694..805F}
{Farihi}, J., {Jura}, M., \& {Zuckerman}, B. 2009, \apj, 694, 805, \dodoi{10.1088/0004-637X/694/2/805}

\bibitem[{{Farihi} {et~al.}(2008{\natexlab{b}}){Farihi}, {Zuckerman}, \& {Becklin}}]{2008ApJ...674..431F}
{Farihi}, J., {Zuckerman}, B., \& {Becklin}, E.~E. 2008{\natexlab{b}}, ApJ, 674, 431, \dodoi{10.1086/521715}

\bibitem[{{Farihi} {et~al.}(2022){Farihi}, {Hermes}, {Marsh}, {Mustill}, {Wyatt}, {Guidry}, {Wilson}, {Redfield}, {Izquierdo}, {Toloza}, {G{\"a}nsicke}, {Aungwerojwit}, {Kaewmanee}, {Dhillon}, \& {Swan}}]{2022MNRAS.511.1647F}
{Farihi}, J., {Hermes}, J.~J., {Marsh}, T.~R., {et~al.} 2022, \mnras, 511, 1647, \dodoi{10.1093/mnras/stab3475}

\bibitem[{Foreman-Mackey(2016)}]{corner}
Foreman-Mackey, D. 2016, The Journal of Open Source Software, 1, 24, \dodoi{10.21105/joss.00024}

\bibitem[{{Fortney} {et~al.}(2007){Fortney}, {Marley}, \& {Barnes}}]{2007ApJ...659.1661F}
{Fortney}, J.~J., {Marley}, M.~S., \& {Barnes}, J.~W. 2007, \apj, 659, 1661, \dodoi{10.1086/512120}

\bibitem[{{Fossati} {et~al.}(2012){Fossati}, {Bagnulo}, {Haswell}, {Patel}, {Busuttil}, {Kowalski}, {Shulyak}, \& {Sterzik}}]{2012ApJ...757L..15F}
{Fossati}, L., {Bagnulo}, S., {Haswell}, C.~A., {et~al.} 2012, ApJL, 757, L15, \dodoi{10.1088/2041-8205/757/1/L15}

\bibitem[{{Gaensicke} {et~al.}(2015){Gaensicke}, {Debes}, {Dufour}, {Farihi}, {Gentile Fusillo}, {Hermes}, {Holberg}, {Koester}, {Kowalski}, {Subasavage}, {Toonen}, \& {Tremblay}}]{2015hst..prop14076G}
{Gaensicke}, B.~T., {Debes}, J.~H., {Dufour}, P., {et~al.} 2015, {An HST legacy ultraviolet spectroscopic survey of the 13pc white dwarf sample}, HST Proposal. Cycle 23, ID. \#14076

\bibitem[{{Gaia Collaboration}(2020)}]{2020yCat.1350....0G}
{Gaia Collaboration}. 2020, {VizieR Online Data Catalog: Gaia EDR3 (Gaia Collaboration, 2020)}, VizieR On-line Data Catalog: I/350. Originally published in: 2021A\&A...649A...1G; doi:10.5270/esa-1ug, \dodoi{10.26093/cds/vizier.1350}

\bibitem[{{Gaia Collaboration} {et~al.}(2016){Gaia Collaboration}, {Prusti}, {de Bruijne}, {Brown}, {Vallenari}, {Babusiaux}, {Bailer-Jones}, {Bastian}, {Biermann}, {Evans}, {Eyer}, {Jansen}, {Jordi}, {Klioner}, {Lammers}, {Lindegren}, {Luri}, {Mignard}, {Milligan}, {Panem}, {Poinsignon}, {Pourbaix}, {Randich}, {Sarri}, {Sartoretti}, {Siddiqui}, {Soubiran}, {Valette}, {van Leeuwen}, {Walton}, {Aerts}, {Arenou}, {Cropper}, {Drimmel}, {H{\o}g}, {Katz}, {Lattanzi}, {O'Mullane}, {Grebel}, {Holland}, {Huc}, {Passot}, {Bramante}, {Cacciari}, {Casta{\~n}eda}, {Chaoul}, {Cheek}, {De Angeli}, {Fabricius}, {Guerra}, {Hern{\'a}ndez}, {Jean-Antoine-Piccolo}, {Masana}, {Messineo}, {Mowlavi}, {Nienartowicz}, {Ord{\'o}{\~n}ez-Blanco}, {Panuzzo}, {Portell}, {Richards}, {Riello}, {Seabroke}, {Tanga}, {Th{\'e}venin}, {Torra}, {Els}, {Gracia-Abril}, {Comoretto}, {Garcia-Reinaldos}, {Lock}, {Mercier}, {Altmann}, {Andrae}, {Astraatmadja}, {Bellas-Velidis}, {Benson}, {Berthier}, {Blomme}, {Busso}, {Carry}, {Cellino}, {Clementini},
  {Cowell}, {Creevey}, {Cuypers}, {Davidson}, {De Ridder}, {de Torres}, {Delchambre}, {Dell'Oro}, {Ducourant}, {Fr{\'e}mat}, {Garc{\'\i}a-Torres}, {Gosset}, {Halbwachs}, {Hambly}, {Harrison}, {Hauser}, {Hestroffer}, {Hodgkin}, {Huckle}, {Hutton}, {Jasniewicz}, {Jordan}, {Kontizas}, {Korn}, {Lanzafame}, {Manteiga}, {Moitinho}, {Muinonen}, {Osinde}, {Pancino}, {Pauwels}, {Petit}, {Recio-Blanco}, {Robin}, {Sarro}, {Siopis}, {Smith}, {Smith}, {Sozzetti}, {Thuillot}, {van Reeven}, {Viala}, {Abbas}, {Abreu Aramburu}, {Accart}, {Aguado}, {Allan}, {Allasia}, {Altavilla}, {{\'A}lvarez}, {Alves}, {Anderson}, {Andrei}, {Anglada Varela}, {Antiche}, {Antoja}, {Ant{\'o}n}, {Arcay}, {Atzei}, {Ayache}, {Bach}, {Baker}, {Balaguer-N{\'u}{\~n}ez}, {Barache}, {Barata}, {Barbier}, {Barblan}, {Baroni}, {Barrado y Navascu{\'e}s}, {Barros}, {Barstow}, {Becciani}, {Bellazzini}, {Bellei}, {Bello Garc{\'\i}a}, {Belokurov}, {Bendjoya}, {Berihuete}, {Bianchi}, {Bienaym{\'e}}, {Billebaud}, {Blagorodnova}, {Blanco-Cuaresma}, {Boch},
  {Bombrun}, {Borrachero}, {Bouquillon}, {Bourda}, {Bouy}, {Bragaglia}, {Breddels}, {Brouillet}, {Br{\"u}semeister}, {Bucciarelli}, {Budnik}, {Burgess}, {Burgon}, {Burlacu}, {Busonero}, {Buzzi}, {Caffau}, {Cambras}, {Campbell}, {Cancelliere}, {Cantat-Gaudin}, {Carlucci}, {Carrasco}, {Castellani}, {Charlot}, {Charnas}, {Charvet}, {Chassat}, {Chiavassa}, {Clotet}, {Cocozza}, {Collins}, {Collins}, {Costigan}, {Crifo}, {Cross}, {Crosta}, {Crowley}, {Dafonte}, {Damerdji}, {Dapergolas}, {David}, {David}, {De Cat}, {de Felice}, {de Laverny}, {De Luise}, {De March}, {de Martino}, {de Souza}, {Debosscher}, {del Pozo}, {Delbo}, {Delgado}, {Delgado}, {di Marco}, {Di Matteo}, {Diakite}, {Distefano}, {Dolding}, {Dos Anjos}, {Drazinos}, {Dur{\'a}n}, {Dzigan}, {Ecale}, {Edvardsson}, {Enke}, {Erdmann}, {Escolar}, {Espina}, {Evans}, {Eynard Bontemps}, {Fabre}, {Fabrizio}, {Faigler}, {Falc{\~a}o}, {Farr{\`a}s Casas}, {Faye}, {Federici}, {Fedorets}, {Fern{\'a}ndez-Hern{\'a}ndez}, {Fernique}, {Fienga}, {Figueras}, {Filippi},
  {Findeisen}, {Fonti}, {Fouesneau}, {Fraile}, {Fraser}, {Fuchs}, {Furnell}, {Gai}, {Galleti}, {Galluccio}, {Garabato}, {Garc{\'\i}a-Sedano}, {Gar{\'e}}, {Garofalo}, {Garralda}, {Gavras}, {Gerssen}, {Geyer}, {Gilmore}, {Girona}, {Giuffrida}, {Gomes}, {Gonz{\'a}lez-Marcos}, {Gonz{\'a}lez-N{\'u}{\~n}ez}, {Gonz{\'a}lez-Vidal}, {Granvik}, {Guerrier}, {Guillout}, {Guiraud}, {G{\'u}rpide}, {Guti{\'e}rrez-S{\'a}nchez}, {Guy}, {Haigron}, {Hatzidimitriou}, {Haywood}, {Heiter}, {Helmi}, {Hobbs}, {Hofmann}, {Holl}, {Holland}, {Hunt}, {Hypki}, {Icardi}, {Irwin}, {Jevardat de Fombelle}, {Jofr{\'e}}, {Jonker}, {Jorissen}, {Julbe}, {Karampelas}, {Kochoska}, {Kohley}, {Kolenberg}, {Kontizas}, {Koposov}, {Kordopatis}, {Koubsky}, {Kowalczyk}, {Krone-Martins}, {Kudryashova}, {Kull}, {Bachchan}, {Lacoste-Seris}, {Lanza}, {Lavigne}, {Le Poncin-Lafitte}, {Lebreton}, {Lebzelter}, {Leccia}, {Leclerc}, {Lecoeur-Taibi}, {Lemaitre}, {Lenhardt}, {Leroux}, {Liao}, {Licata}, {Lindstr{\o}m}, {Lister}, {Livanou}, {Lobel}, {L{\"o}ffler},
  {L{\'o}pez}, {Lopez-Lozano}, {Lorenz}, {Loureiro}, {MacDonald}, {Magalh{\~a}es Fernandes}, {Managau}, {Mann}, {Mantelet}, {Marchal}, {Marchant}, {Marconi}, {Marie}, {Marinoni}, {Marrese}, {Marschalk{\'o}}, {Marshall}, {Mart{\'\i}n-Fleitas}, {Martino}, {Mary}, {Matijevi{\v{c}}}, {Mazeh}, {McMillan}, {Messina}, {Mestre}, {Michalik}, {Millar}, {Miranda}, {Molina}, {Molinaro}, {Molinaro}, {Moln{\'a}r}, {Moniez}, {Montegriffo}, {Monteiro}, {Mor}, {Mora}, {Morbidelli}, {Morel}, {Morgenthaler}, {Morley}, {Morris}, {Mulone}, {Muraveva}, {Musella}, {Narbonne}, {Nelemans}, {Nicastro}, {Noval}, {Ord{\'e}novic}, {Ordieres-Mer{\'e}}, {Osborne}, {Pagani}, {Pagano}, {Pailler}, {Palacin}, {Palaversa}, {Parsons}, {Paulsen}, {Pecoraro}, {Pedrosa}, {Pentik{\"a}inen}, {Pereira}, {Pichon}, {Piersimoni}, {Pineau}, {Plachy}, {Plum}, {Poujoulet}, {Pr{\v{s}}a}, {Pulone}, {Ragaini}, {Rago}, {Rambaux}, {Ramos-Lerate}, {Ranalli}, {Rauw}, {Read}, {Regibo}, {Renk}, {Reyl{\'e}}, {Ribeiro}, {Rimoldini}, {Ripepi}, {Riva}, {Rixon},
  {Roelens}, {Romero-G{\'o}mez}, {Rowell}, {Royer}, {Rudolph}, {Ruiz-Dern}, {Sadowski}, {Sagrist{\`a} Sell{\'e}s}, {Sahlmann}, {Salgado}, {Salguero}, {Sarasso}, {Savietto}, {Schnorhk}, {Schultheis}, {Sciacca}, {Segol}, {Segovia}, {Segransan}, {Serpell}, {Shih}, {Smareglia}, {Smart}, {Smith}, {Solano}, {Solitro}, {Sordo}, {Soria Nieto}, {Souchay}, {Spagna}, {Spoto}, {Stampa}, {Steele}, {Steidelm{\"u}ller}, {Stephenson}, {Stoev}, {Suess}, {S{\"u}veges}, {Surdej}, {Szabados}, {Szegedi-Elek}, {Tapiador}, {Taris}, {Tauran}, {Taylor}, {Teixeira}, {Terrett}, {Tingley}, {Trager}, {Turon}, {Ulla}, {Utrilla}, {Valentini}, {van Elteren}, {Van Hemelryck}, {van Leeuwen}, {Varadi}, {Vecchiato}, {Veljanoski}, {Via}, {Vicente}, {Vogt}, {Voss}, {Votruba}, {Voutsinas}, {Walmsley}, {Weiler}, {Weingrill}, {Werner}, {Wevers}, {Whitehead}, {Wyrzykowski}, {Yoldas}, {{\v{Z}}erjal}, {Zucker}, {Zurbach}, {Zwitter}, {Alecu}, {Allen}, {Allende Prieto}, {Amorim}, {Anglada-Escud{\'e}}, {Arsenijevic}, {Azaz}, {Balm}, {Beck}, {Bernstein},
  {Bigot}, {Bijaoui}, {Blasco}, {Bonfigli}, {Bono}, {Boudreault}, {Bressan}, {Brown}, {Brunet}, {Bunclark}, {Buonanno}, {Butkevich}, {Carret}, {Carrion}, {Chemin}, {Ch{\'e}reau}, {Corcione}, {Darmigny}, {de Boer}, {de Teodoro}, {de Zeeuw}, {Delle Luche}, {Domingues}, {Dubath}, {Fodor}, {Fr{\'e}zouls}, {Fries}, {Fustes}, {Fyfe}, {Gallardo}, {Gallegos}, {Gardiol}, {Gebran}, {Gomboc}, {G{\'o}mez}, {Grux}, {Gueguen}, {Heyrovsky}, {Hoar}, {Iannicola}, {Isasi Parache}, {Janotto}, {Joliet}, {Jonckheere}, {Keil}, {Kim}, {Klagyivik}, {Klar}, {Knude}, {Kochukhov}, {Kolka}, {Kos}, {Kutka}, {Lainey}, {LeBouquin}, {Liu}, {Loreggia}, {Makarov}, {Marseille}, {Martayan}, {Martinez-Rubi}, {Massart}, {Meynadier}, {Mignot}, {Munari}, {Nguyen}, {Nordlander}, {Ocvirk}, {O'Flaherty}, {Olias Sanz}, {Ortiz}, {Osorio}, {Oszkiewicz}, {Ouzounis}, {Palmer}, {Park}, {Pasquato}, {Peltzer}, {Peralta}, {P{\'e}turaud}, {Pieniluoma}, {Pigozzi}, {Poels}, {Prat}, {Prod'homme}, {Raison}, {Rebordao}, {Risquez}, {Rocca-Volmerange}, {Rosen},
  {Ruiz-Fuertes}, {Russo}, {Sembay}, {Serraller Vizcaino}, {Short}, {Siebert}, {Silva}, {Sinachopoulos}, {Slezak}, {Soffel}, {Sosnowska}, {Strai{\v{z}}ys}, {ter Linden}, {Terrell}, {Theil}, {Tiede}, {Troisi}, {Tsalmantza}, {Tur}, {Vaccari}, {Vachier}, {Valles}, {Van Hamme}, {Veltz}, {Virtanen}, {Wallut}, {Wichmann}, {Wilkinson}, {Ziaeepour}, \& {Zschocke}}]{Gaia}
{Gaia Collaboration}, {Prusti}, T., {de Bruijne}, J.~H.~J., {et~al.} 2016, \aap, 595, A1, \dodoi{10.1051/0004-6361/201629272}

\bibitem[{{G{\"a}nsicke} {et~al.}(2006){G{\"a}nsicke}, {Marsh}, {Southworth}, \& {Rebassa-Mansergas}}]{2006Sci...314.1908G}
{G{\"a}nsicke}, B.~T., {Marsh}, T.~R., {Southworth}, J., \& {Rebassa-Mansergas}, A. 2006, Science, 314, 1908, \dodoi{10.1126/science.1135033}

\bibitem[{{G{\"a}nsicke} {et~al.}(2019){G{\"a}nsicke}, {Schreiber}, {Toloza}, {Gentile Fusillo}, {Koester}, \& {Manser}}]{2019Natur.576...61G}
{G{\"a}nsicke}, B.~T., {Schreiber}, M.~R., {Toloza}, O., {et~al.} 2019, Nature, 576, 61, \dodoi{10.1038/s41586-019-1789-8}

\bibitem[{{Gentile Fusillo} {et~al.}(2021){Gentile Fusillo}, {Tremblay}, {Cukanovaite}, {Vorontseva}, {Lallement}, {Hollands}, {G{\"a}nsicke}, {Burdge}, {McCleery}, \& {Jordan}}]{2021MNRAS.508.3877G}
{Gentile Fusillo}, N.~P., {Tremblay}, P.~E., {Cukanovaite}, E., {et~al.} 2021, MNRAS, 508, 3877, \dodoi{10.1093/mnras/stab2672}

\bibitem[{{Gillon} {et~al.}(2016){Gillon}, {Jehin}, {Lederer}, {Delrez}, {de Wit}, {Burdanov}, {Van Grootel}, {Burgasser}, {Triaud}, {Opitom}, {Demory}, {Sahu}, {Bardalez Gagliuffi}, {Magain}, \& {Queloz}}]{2016Natur.533..221G}
{Gillon}, M., {Jehin}, E., {Lederer}, S.~M., {et~al.} 2016, \nat, 533, 221, \dodoi{10.1038/nature17448}

\bibitem[{{Girven} {et~al.}(2011){Girven}, {G{\"a}nsicke}, {Steeghs}, \& {Koester}}]{2011MNRAS.417.1210G}
{Girven}, J., {G{\"a}nsicke}, B.~T., {Steeghs}, D., \& {Koester}, D. 2011, MNRAS, 417, 1210, \dodoi{10.1111/j.1365-2966.2011.19337.x}

\bibitem[{{Gomez Gonzalez} {et~al.}(2017){Gomez Gonzalez}, {Wertz}, {Absil}, {Christiaens}, {Defr{\`e}re}, {Mawet}, {Milli}, {Absil}, {Van Droogenbroeck}, {Cantalloube}, {Hinz}, {Skemer}, {Karlsson}, \& {Surdej}}]{VIP}
{Gomez Gonzalez}, C.~A., {Wertz}, O., {Absil}, O., {et~al.} 2017, \aj, 154, 7, \dodoi{10.3847/1538-3881/aa73d7}

\bibitem[{{Gordon} {et~al.}(2022){Gordon}, {Bohlin}, {Sloan}, {Rieke}, {Volk}, {Boyer}, {Muzerolle}, {Schlawin}, {Deustua}, {Hines}, {Kraemer}, {Mullally}, \& {Su}}]{2022arXiv220406500G}
{Gordon}, K.~D., {Bohlin}, R., {Sloan}, G.~C., {et~al.} 2022, arXiv e-prints, arXiv:2204.06500.
\newblock \doarXiv{2204.06500}

\bibitem[{{Hogg} {et~al.}(2020){Hogg}, {Casewell}, {Wynn}, {Longstaff}, {Braker}, {Burleigh}, {Tilbrook}, {Geier}, {Koester}, {Debes}, \& {Lodieu}}]{2020MNRAS.498...12H}
{Hogg}, M.~A., {Casewell}, S.~L., {Wynn}, G.~A., {et~al.} 2020, MNRAS, 498, 12, \dodoi{10.1093/mnras/staa2233}

\bibitem[{{Holberg} {et~al.}(2008){Holberg}, {Sion}, {Oswalt}, {McCook}, {Foran}, \& {Subasavage}}]{Holberg2008}
{Holberg}, J.~B., {Sion}, E.~M., {Oswalt}, T., {et~al.} 2008, \aj, 135, 1225, \dodoi{10.1088/0004-6256/135/4/1225}

\bibitem[{{Janson} {et~al.}(2021{\natexlab{a}}){Janson}, {Gratton}, {Rodet}, {Vigan}, {Bonnefoy}, {Delorme}, {Mamajek}, {Reffert}, {Stock}, {Marleau}, {Langlois}, {Chauvin}, {Desidera}, {Ringqvist}, {Mayer}, {Viswanath}, {Squicciarini}, {Meyer}, {Samland}, {Petrus}, {Helled}, {Kenworthy}, {Quanz}, {Biller}, {Henning}, {Mesa}, {Engler}, \& {Carson}}]{Janson2021bcen}
{Janson}, M., {Gratton}, R., {Rodet}, L., {et~al.} 2021{\natexlab{a}}, \nat, 600, 231, \dodoi{10.1038/s41586-021-04124-8}

\bibitem[{{Janson} {et~al.}(2021{\natexlab{b}}){Janson}, {Squicciarini}, {Delorme}, {Gratton}, {Bonnefoy}, {Reffert}, {Mamajek}, {Eriksson}, {Vigan}, {Langlois}, {Engler}, {Chauvin}, {Desidera}, {Mayer}, {Marleau}, {Bohn}, {Samland}, {Meyer}, {d'Orazi}, {Henning}, {Quanz}, {Kenworthy}, \& {Carson}}]{2021A&A...646A.164J}
{Janson}, M., {Squicciarini}, V., {Delorme}, P., {et~al.} 2021{\natexlab{b}}, \aap, 646, A164, \dodoi{10.1051/0004-6361/202039683}

\bibitem[{{Johnson} {et~al.}(2013){Johnson}, {Morton}, \& {Wright}}]{Johnson2013}
{Johnson}, J.~A., {Morton}, T.~D., \& {Wright}, J.~T. 2013, \apj, 763, 53, \dodoi{10.1088/0004-637X/763/1/53}

\bibitem[{{Johnson} {et~al.}(2007){Johnson}, {Fischer}, {Marcy}, {Wright}, {Driscoll}, {Butler}, {Hekker}, {Reffert}, \& {Vogt}}]{Johnson2007}
{Johnson}, J.~A., {Fischer}, D.~A., {Marcy}, G.~W., {et~al.} 2007, \apj, 665, 785, \dodoi{10.1086/519677}

\bibitem[{{Jura}(2003)}]{2003ApJ...584L..91J}
{Jura}, M. 2003, \apjl, 584, L91, \dodoi{10.1086/374036}

\bibitem[{{Jura} {et~al.}(2007){Jura}, {Farihi}, \& {Zuckerman}}]{2007ApJ...663.1285J}
{Jura}, M., {Farihi}, J., \& {Zuckerman}, B. 2007, \apj, 663, 1285, \dodoi{10.1086/518767}

\bibitem[{{Jura} {et~al.}(2009){Jura}, {Farihi}, \& {Zuckerman}}]{2009AJ....137.3191J}
---. 2009, \aj, 137, 3191, \dodoi{10.1088/0004-6256/137/2/3191}

\bibitem[{{Kervella} {et~al.}(2019){Kervella}, {Arenou}, {Mignard}, \& {Th{\'e}venin}}]{Kervella2019}
{Kervella}, P., {Arenou}, F., {Mignard}, F., \& {Th{\'e}venin}, F. 2019, \aap, 623, A72, \dodoi{10.1051/0004-6361/201834371}

\bibitem[{{Kilic} {et~al.}(2010){Kilic}, {Brown}, \& {McLeod}}]{2010ApJ...708..411K}
{Kilic}, M., {Brown}, W.~R., \& {McLeod}, B. 2010, ApJ, 708, 411, \dodoi{10.1088/0004-637X/708/1/411}

\bibitem[{{Kilic} {et~al.}(2005){Kilic}, {von Hippel}, {Leggett}, \& {Winget}}]{2005ApJ...632L.115K}
{Kilic}, M., {von Hippel}, T., {Leggett}, S.~K., \& {Winget}, D.~E. 2005, ApJL, 632, L115, \dodoi{10.1086/497825}

\bibitem[{{Kilic} {et~al.}(2006){Kilic}, {von Hippel}, {Leggett}, \& {Winget}}]{2006ApJ...646..474K}
---. 2006, \apj, 646, 474, \dodoi{10.1086/504682}

\bibitem[{{Kiman} {et~al.}(2022){Kiman}, {Xu}, {Faherty}, {Gagn{\'e}}, {Angus}, {Brandt}, {Casewell}, \& {Cruz}}]{2022AJ....164...62K}
{Kiman}, R., {Xu}, S., {Faherty}, J.~K., {et~al.} 2022, \aj, 164, 62, \dodoi{10.3847/1538-3881/ac7788}

\bibitem[{{Klein} {et~al.}(2010){Klein}, {Jura}, {Koester}, {Zuckerman}, \& {Melis}}]{2010ApJ...709..950K}
{Klein}, B., {Jura}, M., {Koester}, D., {Zuckerman}, B., \& {Melis}, C. 2010, \apj, 709, 950, \dodoi{10.1088/0004-637X/709/2/950}

\bibitem[{{Koen} {et~al.}(2010){Koen}, {Kilkenny}, {van Wyk}, \& {Marang}}]{2010MNRAS.403.1949K}
{Koen}, C., {Kilkenny}, D., {van Wyk}, F., \& {Marang}, F. 2010, \mnras, 403, 1949, \dodoi{10.1111/j.1365-2966.2009.16182.x}

\bibitem[{{Koester}(2009)}]{2009A&A...498..517K}
{Koester}, D. 2009, \aap, 498, 517, \dodoi{10.1051/0004-6361/200811468}

\bibitem[{{Koester} {et~al.}(2014){Koester}, {G{\"a}nsicke}, \& {Farihi}}]{2014A&A...566A..34K}
{Koester}, D., {G{\"a}nsicke}, B.~T., \& {Farihi}, J. 2014, \aap, 566, A34, \dodoi{10.1051/0004-6361/201423691}

\bibitem[{{Koester} {et~al.}(2009){Koester}, {Voss}, {Napiwotzki}, {Christlieb}, {Homeier}, {Lisker}, {Reimers}, \& {Heber}}]{2009A&A...505..441K}
{Koester}, D., {Voss}, B., {Napiwotzki}, R., {et~al.} 2009, \aap, 505, 441, \dodoi{10.1051/0004-6361/200912531}

\bibitem[{{Koester} \& {Wilken}(2006)}]{2006A&A...453.1051K}
{Koester}, D., \& {Wilken}, D. 2006, \aap, 453, 1051, \dodoi{10.1051/0004-6361:20064843}

\bibitem[{{Korol} {et~al.}(2022){Korol}, {Belokurov}, \& {Toonen}}]{Korol22}
{Korol}, V., {Belokurov}, V., \& {Toonen}, S. 2022, \mnras, 515, 1228, \dodoi{10.1093/mnras/stac1686}

\bibitem[{{Kozakis} {et~al.}(2018){Kozakis}, {Kaltenegger}, \& {Hoard}}]{2018ApJ...862...69K}
{Kozakis}, T., {Kaltenegger}, L., \& {Hoard}, D.~W. 2018, ApJ, 862, 69, \dodoi{10.3847/1538-4357/aacbc7}

\bibitem[{{Kozakis} {et~al.}(2020){Kozakis}, {Lin}, \& {Kaltenegger}}]{2020ApJ...894L...6K}
{Kozakis}, T., {Lin}, Z., \& {Kaltenegger}, L. 2020, ApJL, 894, L6, \dodoi{10.3847/2041-8213/ab6f6a}

\bibitem[{{Lagos} {et~al.}(2021){Lagos}, {Schreiber}, {Zorotovic}, {G{\"a}nsicke}, {Ronco}, \& {Hamers}}]{2021MNRAS.501..676L}
{Lagos}, F., {Schreiber}, M.~R., {Zorotovic}, M., {et~al.} 2021, MNRAS, 501, 676, \dodoi{10.1093/mnras/staa3703}

\bibitem[{{Lagrange} {et~al.}(2010){Lagrange}, {Bonnefoy}, {Chauvin}, {Apai}, {Ehrenreich}, {Boccaletti}, {Gratadour}, {Rouan}, {Mouillet}, {Lacour}, \& {Kasper}}]{Lagrange2010}
{Lagrange}, A.~M., {Bonnefoy}, M., {Chauvin}, G., {et~al.} 2010, Science, 329, 57, \dodoi{10.1126/science.1187187}

\bibitem[{Lai {et~al.}(2021)Lai, Dennihy, Xu, Nitta, Kleinman, Leggett, Bonsor, Hodgkin, Rebassa-Mansergas, \& Rogers}]{lai2021infrared}
Lai, S., Dennihy, E., Xu, S., {et~al.} 2021, Infrared Excesses around Bright White Dwarfs from Gaia and unWISE. II.
\newblock \doarXiv{2107.01221}

\bibitem[{{Ledda} {et~al.}(2023){Ledda}, {Danielski}, \& {Turrini}}]{2023A&A...675A.184L}
{Ledda}, S., {Danielski}, C., \& {Turrini}, D. 2023, \aap, 675, A184, \dodoi{10.1051/0004-6361/202245827}

\bibitem[{{Libby-Roberts} {et~al.}(2020){Libby-Roberts}, {Berta-Thompson}, {D{\'e}sert}, {Masuda}, {Morley}, {Lopez}, {Deck}, {Fabrycky}, {Fortney}, {Line}, {Sanchis-Ojeda}, \& {Winn}}]{2020AJ....159...57L}
{Libby-Roberts}, J.~E., {Berta-Thompson}, Z.~K., {D{\'e}sert}, J.-M., {et~al.} 2020, \aj, 159, 57, \dodoi{10.3847/1538-3881/ab5d36}

\bibitem[{{Limbach} {et~al.}(2022){Limbach}, {Vanderburg}, {Stevenson}, {Blouin}, {Morley}, {Lustig-Yaeger}, {Soares-Furtado}, \& {Janson}}]{Limbach22}
{Limbach}, M.~A., {Vanderburg}, A., {Stevenson}, K.~B., {et~al.} 2022, \mnras, 517, 2622, \dodoi{10.1093/mnras/stac2823}

\bibitem[{{Limbach} {et~al.}(2023){Limbach}, {Vanderburg}, {Blouin}, {Janson}, {Kenworthy}, {Kleisioti}, {Stevenson}, {Venner}, \& {Morley}}]{2023jwst.prop.4403M}
{Limbach}, M.~A., {Vanderburg}, A., {Blouin}, S., {et~al.} 2023, {The MIRI survey for Exoplanets Orbiting White-dwarfs (MEOW)}, JWST Proposal. Cycle 2, ID. \#3964

\bibitem[{{Limbach} {et~al.}(2024){Limbach}, {MacDonald}, {Vanderburg}, {Blouin}, {Gallo}, {Jenkins}, {Morley}, {O'Connor}, {Stevenson}, \& {Xu}}]{2024jwst.prop.5204L}
{Limbach}, M.~A., {MacDonald}, R., {Vanderburg}, A., {et~al.} 2024, {Probing the Dynamical History and the Mid-IR SED of WD 1856b}, JWST Proposal. Cycle 3, ID. \#5204

\bibitem[{Lindegren(2018)}]{Lindegren18}
Lindegren, L. 2018, Re-normalising the astrometric chi-square in Gaia DR2, GAIA-C3-TN-LU-LL-124-01.
\newblock \url{http://www.rssd.esa.int/doc_fetch.php?id=3757412}

\bibitem[{{Lindegren} {et~al.}(2021){Lindegren}, {Klioner}, {Hern{\'a}ndez}, {Bombrun}, {Ramos-Lerate}, {Steidelm{\"u}ller}, {Bastian}, {Biermann}, {de Torres}, {Gerlach}, {Geyer}, {Hilger}, {Hobbs}, {Lammers}, {McMillan}, {Stephenson}, {Casta{\~n}eda}, {Davidson}, {Fabricius}, {Gracia-Abril}, {Portell}, {Rowell}, {Teyssier}, {Torra}, {Bartolom{\'e}}, {Clotet}, {Garralda}, {Gonz{\'a}lez-Vidal}, {Torra}, {Abbas}, {Altmann}, {Anglada Varela}, {Balaguer-N{\'u}{\~n}ez}, {Balog}, {Barache}, {Becciani}, {Bernet}, {Bertone}, {Bianchi}, {Bouquillon}, {Brown}, {Bucciarelli}, {Busonero}, {Butkevich}, {Buzzi}, {Cancelliere}, {Carlucci}, {Charlot}, {Cioni}, {Crosta}, {Crowley}, {del Peloso}, {del Pozo}, {Drimmel}, {Esquej}, {Fienga}, {Fraile}, {Gai}, {Garcia-Reinaldos}, {Guerra}, {Hambly}, {Hauser}, {Jan{\ss}en}, {Jordan}, {Kostrzewa-Rutkowska}, {Lattanzi}, {Liao}, {Licata}, {Lister}, {L{\"o}ffler}, {Marchant}, {Masip}, {Mignard}, {Mints}, {Molina}, {Mora}, {Morbidelli}, {Murphy}, {Pagani}, {Panuzzo}, {Pe{\~n}alosa
  Esteller}, {Poggio}, {Re Fiorentin}, {Riva}, {Sagrist{\`a} Sell{\'e}s}, {Sanchez Gimenez}, {Sarasso}, {Sciacca}, {Siddiqui}, {Smart}, {Souami}, {Spagna}, {Steele}, {Taris}, {Utrilla}, {van Reeven}, \& {Vecchiato}}]{Lindegren21}
{Lindegren}, L., {Klioner}, S.~A., {Hern{\'a}ndez}, J., {et~al.} 2021, \aap, 649, A2, \dodoi{10.1051/0004-6361/20203970910.48550/arXiv.2012.03380}

\bibitem[{{Linder} {et~al.}(2019){Linder}, {Mordasini}, {Molli{\`e}re}, {Marleau}, {Malik}, {Quanz}, \& {Meyer}}]{2019A&A...623A..85L}
{Linder}, E.~F., {Mordasini}, C., {Molli{\`e}re}, P., {et~al.} 2019, AAP, 623, A85, \dodoi{10.1051/0004-6361/201833873}

\bibitem[{{Lloyd}(2011)}]{Lloyd2011}
{Lloyd}, J.~P. 2011, \apjl, 739, L49, \dodoi{10.1088/2041-8205/739/2/L49}

\bibitem[{{Loeb} \& {Maoz}(2013)}]{2013MNRAS.432L..11L}
{Loeb}, A., \& {Maoz}, D. 2013, MNRAS, 432, L11, \dodoi{10.1093/mnrasl/slt026}

\bibitem[{{Luhman} {et~al.}(2011){Luhman}, {Burgasser}, \& {Bochanski}}]{2011ApJ...730L...9L}
{Luhman}, K.~L., {Burgasser}, A.~J., \& {Bochanski}, J.~J. 2011, ApJL, 730, L9, \dodoi{10.1088/2041-8205/730/1/L9}

\bibitem[{{Luhman} {et~al.}(2024){Luhman}, {Tremblin}, {Alves de Oliveira}, {Birkmann}, {Baraffe}, {Chabrier}, {Manjavacas}, {Parker}, \& {Valenti}}]{2024AJ....167....5L}
{Luhman}, K.~L., {Tremblin}, P., {Alves de Oliveira}, C., {et~al.} 2024, \aj, 167, 5, \dodoi{10.3847/1538-3881/ad0b72}

\bibitem[{{MacDonald}(2023)}]{2023JOSS....8.4873M}
{MacDonald}, R.~J. 2023, The Journal of Open Source Software, 8, 4873, \dodoi{10.21105/joss.04873}

\bibitem[{{MacDonald} \& {Madhusudhan}(2017)}]{2017MNRAS.469.1979M}
{MacDonald}, R.~J., \& {Madhusudhan}, N. 2017, \mnras, 469, 1979, \dodoi{10.1093/mnras/stx804}

\bibitem[{{MacDonald} {et~al.}(2021){MacDonald}, {Batalha}, {Foote}, {Kozakis}, {Lewis}, {Lothringer}, {Marley}, {May}, {Mayorga}, {Mishra}, {Mullally}, {O'Connor}, {Sing}, {Valenti}, \& {Vanderburg}}]{2021jwst.prop.2358M}
{MacDonald}, R.~J., {Batalha}, N., {Foote}, T.~O., {et~al.} 2021, {Under the Light of a Dead Star: Revealing the Atmospheric Composition of a White Dwarf Planet}, JWST Proposal. Cycle 1, ID. \#2358

\bibitem[{{Maldonado} {et~al.}(2022){Maldonado}, {Villaver}, {Mustill}, \& {Ch{\'a}vez}}]{2022MNRAS.512..104M}
{Maldonado}, R.~F., {Villaver}, E., {Mustill}, A.~J., \& {Ch{\'a}vez}, M. 2022, \mnras, 512, 104, \dodoi{10.1093/mnras/stac481}

\bibitem[{{Maldonado} {et~al.}(2021){Maldonado}, {Villaver}, {Mustill}, {Ch{\'a}vez}, \& {Bertone}}]{2021MNRAS.501L..43M}
{Maldonado}, R.~F., {Villaver}, E., {Mustill}, A.~J., {Ch{\'a}vez}, M., \& {Bertone}, E. 2021, MNRAS, 501, L43, \dodoi{10.1093/mnrasl/slaa193}

\bibitem[{{Marley} {et~al.}(2021){Marley}, {Saumon}, {Visscher}, {Lupu}, {Freedman}, {Morley}, {Fortney}, {Seay}, {Smith}, {Teal}, \& {Wang}}]{Marley2021}
{Marley}, M.~S., {Saumon}, D., {Visscher}, C., {et~al.} 2021, ApJ, 920, 85, \dodoi{10.3847/1538-4357/ac141d}

\bibitem[{{Masuda}(2014)}]{2014ApJ...783...53M}
{Masuda}, K. 2014, \apj, 783, 53, \dodoi{10.1088/0004-637X/783/1/53}

\bibitem[{{Melis} {et~al.}(2011){Melis}, {Farihi}, {Dufour}, {Zuckerman}, {Burgasser}, {Bergeron}, {Bochanski}, \& {Simcoe}}]{2011ApJ...732...90M}
{Melis}, C., {Farihi}, J., {Dufour}, P., {et~al.} 2011, \apj, 732, 90, \dodoi{10.1088/0004-637X/732/2/90}

\bibitem[{{Merlov} {et~al.}(2021){Merlov}, {Bear}, \& {Soker}}]{Merlov2021ApJL}
{Merlov}, A., {Bear}, E., \& {Soker}, N. 2021, \apjl, 915, L34, \dodoi{10.3847/2041-8213/ac0f7d}

\bibitem[{{Mu{\~n}oz} \& {Petrovich}(2020)}]{2020ApJ...904L...3M}
{Mu{\~n}oz}, D.~J., \& {Petrovich}, C. 2020, ApJL, 904, L3, \dodoi{10.3847/2041-8213/abc564}

\bibitem[{{Mullally} {et~al.}(2007){Mullally}, {Kilic}, {Reach}, {Kuchner}, {von Hippel}, {Burrows}, \& {Winget}}]{2007ApJS..171..206M}
{Mullally}, F., {Kilic}, M., {Reach}, W.~T., {et~al.} 2007, \apjs, 171, 206, \dodoi{10.1086/511858}

\bibitem[{{Mullally} {et~al.}(2024{\natexlab{a}}){Mullally}, {Albert}, {Mullally}, {Barclay}, {Cracraft}, {Debes}, {Hermes}, {Kilic}, {Poulsen}, {Quintana}, {Reach}, \& {Thibault}}]{2024jwst.prop.4857M}
{Mullally}, F., {Albert}, L., {Mullally}, S.~E., {et~al.} 2024{\natexlab{a}}, {Confirmation of Planetary Companions to White Dwarf Stars}, JWST Proposal. Cycle 3, ID. \#4857

\bibitem[{{Mullally} {et~al.}(2021){Mullally}, {Mullally}, {Albert}, {Barclay}, {Debes}, {Kilic}, {Kuchner}, {Quintana}, \& {Reach}}]{2021jwst.prop.1911M}
{Mullally}, S.~E., {Mullally}, F., {Albert}, L., {et~al.} 2021, {A Search for the Giant Planets that Drive White Dwarf Accretion}, JWST Proposal. Cycle 1, ID. \#1911

\bibitem[{{Mullally} {et~al.}(2024{\natexlab{b}}){Mullally}, {Debes}, {Cracraft}, {Mullally}, {Poulsen}, {Albert}, {Thibault}, {Reach}, {Hermes}, {Barclay}, {Kilic}, \& {Quintana}}]{2024arXiv240113153M}
{Mullally}, S.~E., {Debes}, J., {Cracraft}, M., {et~al.} 2024{\natexlab{b}}, arXiv e-prints, arXiv:2401.13153, \dodoi{10.48550/arXiv.2401.13153}

\bibitem[{{Napiwotzki} {et~al.}(2020){Napiwotzki}, {Karl}, {Lisker}, {Catal{\'a}n}, {Drechsel}, {Heber}, {Homeier}, {Koester}, {Leibundgut}, {Marsh}, {Moehler}, {Nelemans}, {Reimers}, {Renzini}, {Str{\"o}er}, \& {Yungelson}}]{2020A&A...638A.131N}
{Napiwotzki}, R., {Karl}, C.~A., {Lisker}, T., {et~al.} 2020, \aap, 638, A131, \dodoi{10.1051/0004-6361/201629648}

\bibitem[{{Nielsen} {et~al.}(2019){Nielsen}, {De Rosa}, {Macintosh}, {Wang}, {Ruffio}, {Chiang}, {Marley}, {Saumon}, {Savransky}, {Ammons}, {Bailey}, {Barman}, {Blain}, {Bulger}, {Burrows}, {Chilcote}, {Cotten}, {Czekala}, {Doyon}, {Duch{\^e}ne}, {Esposito}, {Fabrycky}, {Fitzgerald}, {Follette}, {Fortney}, {Gerard}, {Goodsell}, {Graham}, {Greenbaum}, {Hibon}, {Hinkley}, {Hirsch}, {Hom}, {Hung}, {Dawson}, {Ingraham}, {Kalas}, {Konopacky}, {Larkin}, {Lee}, {Lin}, {Maire}, {Marchis}, {Marois}, {Metchev}, {Millar-Blanchaer}, {Morzinski}, {Oppenheimer}, {Palmer}, {Patience}, {Perrin}, {Poyneer}, {Pueyo}, {Rafikov}, {Rajan}, {Rameau}, {Rantakyr{\"o}}, {Ren}, {Schneider}, {Sivaramakrishnan}, {Song}, {Soummer}, {Tallis}, {Thomas}, {Ward-Duong}, \& {Wolff}}]{Nielsen2019}
{Nielsen}, E.~L., {De Rosa}, R.~J., {Macintosh}, B., {et~al.} 2019, \aj, 158, 13, \dodoi{10.3847/1538-3881/ab16e9}

\bibitem[{{Nordhaus} \& {Spiegel}(2013)}]{Nordhaus2013MNRAS}
{Nordhaus}, J., \& {Spiegel}, D.~S. 2013, \mnras, 432, 500, \dodoi{10.1093/mnras/stt569}

\bibitem[{{Paxton} {et~al.}(2011){Paxton}, {Bildsten}, {Dotter}, {Herwig}, {Lesaffre}, \& {Timmes}}]{Paxton2011}
{Paxton}, B., {Bildsten}, L., {Dotter}, A., {et~al.} 2011, \apjs, 192, 3, \dodoi{10.1088/0067-0049/192/1/3}

\bibitem[{{Paxton} {et~al.}(2013){Paxton}, {Cantiello}, {Arras}, {Bildsten}, {Brown}, {Dotter}, {Mankovich}, {Montgomery}, {Stello}, {Timmes}, \& {Townsend}}]{Paxton2013}
{Paxton}, B., {Cantiello}, M., {Arras}, P., {et~al.} 2013, \apjs, 208, 4, \dodoi{10.1088/0067-0049/208/1/4}

\bibitem[{{Paxton} {et~al.}(2015){Paxton}, {Marchant}, {Schwab}, {Bauer}, {Bildsten}, {Cantiello}, {Dessart}, {Farmer}, {Hu}, {Langer}, {Townsend}, {Townsley}, \& {Timmes}}]{Paxton2015}
{Paxton}, B., {Marchant}, P., {Schwab}, J., {et~al.} 2015, \apjs, 220, 15, \dodoi{10.1088/0067-0049/220/1/15}

\bibitem[{{Poulsen} {et~al.}(2023){Poulsen}, {Debes}, {Farihi}, {Mullally}, {Barclay}, {Quintana}, {Albert}, {Kilic}, \& {Reach}}]{2023jwst.prop.3964M}
{Poulsen}, S., {Debes}, J., {Farihi}, J., {et~al.} 2023, {The MIRI Excess Around Degenerates Survey}, JWST Proposal. Cycle 2, ID. \#3964

\bibitem[{{Poulsen} {et~al.}(2024){Poulsen}, {Debes}, {Cracraft}, {Mullally}, {Reach}, {Kilic}, {Mullally}, {Albert}, {Thibault}, {Hermes}, {Barclay}, \& {Quintana}}]{2024AJ....167..257P}
{Poulsen}, S., {Debes}, J., {Cracraft}, M., {et~al.} 2024, \aj, 167, 257, \dodoi{10.3847/1538-3881/ad374c}

\bibitem[{{Reach} {et~al.}(2005){Reach}, {Kuchner}, {von Hippel}, {Burrows}, {Mullally}, {Kilic}, \& {Winget}}]{2005ApJ...635L.161R}
{Reach}, W.~T., {Kuchner}, M.~J., {von Hippel}, T., {et~al.} 2005, \apjl, 635, L161, \dodoi{10.1086/499561}

\bibitem[{{Rebassa-Mansergas} {et~al.}(2019){Rebassa-Mansergas}, {Solano}, {Xu}, {Rodrigo}, {Jim{\'e}nez-Esteban}, \& {Torres}}]{2019MNRAS.489.3990R}
{Rebassa-Mansergas}, A., {Solano}, E., {Xu}, S., {et~al.} 2019, MNRAS, 489, 3990, \dodoi{10.1093/mnras/stz2423}

\bibitem[{{Reffert} {et~al.}(2015){Reffert}, {Bergmann}, {Quirrenbach}, {Trifonov}, \& {K{\"u}nstler}}]{Reffert2015}
{Reffert}, S., {Bergmann}, C., {Quirrenbach}, A., {Trifonov}, T., \& {K{\"u}nstler}, A. 2015, \aap, 574, A116, \dodoi{10.1051/0004-6361/201322360}

\bibitem[{{Robert} {et~al.}(2024){Robert}, {Farihi}, {Van Eylen}, {Aungwerojwit}, {G{\"a}nsicke}, {Redfield}, {Dhillon}, {Marsh}, \& {Swan}}]{2024arXiv240721743R}
{Robert}, A., {Farihi}, J., {Van Eylen}, V., {et~al.} 2024, arXiv e-prints, arXiv:2407.21743, \dodoi{10.48550/arXiv.2407.21743}

\bibitem[{{Rogers} {et~al.}(2024){Rogers}, {Debes}, {Anslow}, {Bonsor}, {Casewell}, {Dos Santos}, {Dufour}, {G{\"a}nsicke}, {Gentile Fusillo}, {Koester}, {Nielsen}, {Penoyre}, {Rickman}, {Sahlmann}, {Tremblay}, {Vanderburg}, {Xu}, {Dennihy}, {Farihi}, {Hermes}, {Hodgkin}, {Kilic}, {Kowalski}, {Sanderson}, \& {Toonen}}]{2024MNRAS.527..977R}
{Rogers}, L.~K., {Debes}, J., {Anslow}, R.~J., {et~al.} 2024, \mnras, 527, 977, \dodoi{10.1093/mnras/stad3098}

\bibitem[{{Sanderson} {et~al.}(2022){Sanderson}, {Bonsor}, \& {Mustill}}]{Sanderson2022}
{Sanderson}, H., {Bonsor}, A., \& {Mustill}, A. 2022, MNRAS, 517, 5835, \dodoi{10.1093/mnras/stac2867}

\bibitem[{{Sigurdsson} {et~al.}(2003){Sigurdsson}, {Richer}, {Hansen}, {Stairs}, \& {Thorsett}}]{2003Sci...301..193S}
{Sigurdsson}, S., {Richer}, H.~B., {Hansen}, B.~M., {Stairs}, I.~H., \& {Thorsett}, S.~E. 2003, Science, 301, 193, \dodoi{10.1126/science.1086326}

\bibitem[{{Smithsonian Astrophysical Observatory}(2000)}]{2000ascl.soft03002S}
{Smithsonian Astrophysical Observatory}. 2000, {SAOImage DS9: A utility for displaying astronomical images in the X11 window environment}, Astrophysics Source Code Library, record ascl:0003.002

\bibitem[{{Sozzetti}(2005)}]{Sozzetti05}
{Sozzetti}, A. 2005, \pasp, 117, 1021, \dodoi{10.1086/44448710.48550/arXiv.astro-ph/0507115}

\bibitem[{{Stephan} {et~al.}(2021){Stephan}, {Naoz}, \& {Gaudi}}]{2021ApJ...922....4S}
{Stephan}, A.~P., {Naoz}, S., \& {Gaudi}, B.~S. 2021, ApJ, 922, 4, \dodoi{10.3847/1538-4357/ac22a9}

\bibitem[{{Stevenson}(2020)}]{2020ApJ...898L..35S}
{Stevenson}, K.~B. 2020, ApJL, 898, L35, \dodoi{10.3847/2041-8213/aba68c}

\bibitem[{{Su} {et~al.}(2007){Su}, {Chu}, {Rieke}, {Huggins}, {Gruendl}, {Napiwotzki}, {Rauch}, {Latter}, \& {Volk}}]{2007ApJ...657L..41S}
{Su}, K.~Y.~L., {Chu}, Y.~H., {Rieke}, G.~H., {et~al.} 2007, \apjl, 657, L41, \dodoi{10.1086/513018}

\bibitem[{{Swan} {et~al.}(2024){Swan}, {Farihi}, {Su}, \& {Desch}}]{2024MNRAS.529L..41S}
{Swan}, A., {Farihi}, J., {Su}, K. Y.~L., \& {Desch}, S.~J. 2024, \mnras, 529, L41, \dodoi{10.1093/mnrasl/slad198}

\bibitem[{{Thorsett} {et~al.}(1993){Thorsett}, {Arzoumanian}, \& {Taylor}}]{1993ApJ...412L..33T}
{Thorsett}, S.~E., {Arzoumanian}, Z., \& {Taylor}, J.~H. 1993, ApJL, 412, L33, \dodoi{10.1086/186933}

\bibitem[{{Trierweiler} {et~al.}(2023){Trierweiler}, {Doyle}, \& {Young}}]{2023PSJ.....4..136T}
{Trierweiler}, I.~L., {Doyle}, A.~E., \& {Young}, E.~D. 2023, \psj, 4, 136, \dodoi{10.3847/PSJ/acdef3}

\bibitem[{van~der Walt {et~al.}(2011)van~der Walt, Colbert, \& Varoquaux}]{5725236}
van~der Walt, S., Colbert, S.~C., \& Varoquaux, G. 2011, Computing in Science Engineering, 13, 22, \dodoi{10.1109/MCSE.2011.37}

\bibitem[{{van Sluijs} \& {Van Eylen}(2018)}]{2018MNRAS.474.4603V}
{van Sluijs}, L., \& {Van Eylen}, V. 2018, \mnras, 474, 4603, \dodoi{10.1093/mnras/stx3068}

\bibitem[{{Vanderbosch} {et~al.}(2020){Vanderbosch}, {Hermes}, {Dennihy}, {Dunlap}, {Izquierdo}, {Tremblay}, {Cho}, {G{\"a}nsicke}, {Toloza}, {Bell}, {Montgomery}, \& {Winget}}]{2020ApJ...897..171V}
{Vanderbosch}, Z., {Hermes}, J.~J., {Dennihy}, E., {et~al.} 2020, \apj, 897, 171, \dodoi{10.3847/1538-4357/ab9649}

\bibitem[{{Vanderburg} {et~al.}(2021){Vanderburg}, {Becker}, {Daylan}, {Gao}, {Kreidberg}, {MacDonald}, {Morley}, \& {Xu}}]{2021jwst.prop.2507V}
{Vanderburg}, A., {Becker}, J.~C., {Daylan}, T., {et~al.} 2021, {Thermal Emission from the First Planet Transiting a White Dwarf}, JWST Proposal. Cycle 1, ID. \#2507

\bibitem[{{Vanderburg} {et~al.}(2015){Vanderburg}, {Johnson}, {Rappaport}, {Bieryla}, {Irwin}, {Lewis}, {Kipping}, {Brown}, {Dufour}, {Ciardi}, {Angus}, {Schaefer}, {Latham}, {Charbonneau}, {Beichman}, {Eastman}, {McCrady}, {Wittenmyer}, \& {Wright}}]{2015Natur.526..546V}
{Vanderburg}, A., {Johnson}, J.~A., {Rappaport}, S., {et~al.} 2015, Nature, 526, 546, \dodoi{10.1038/nature15527}

\bibitem[{Vanderburg {et~al.}(2020)Vanderburg, Rappaport, Xu, Crossfield, Becker, Gary, Murgas, Blouin, Kaye, Palle, \& et~al.}]{Vanderburg_2020}
Vanderburg, A., Rappaport, S.~A., Xu, S., {et~al.} 2020, Nature, 585, 363–367, \dodoi{10.1038/s41586-020-2713-y}

\bibitem[{{Vauclair} {et~al.}(1997){Vauclair}, {Schmidt}, {Koester}, \& {Allard}}]{Vauclair1997}
{Vauclair}, G., {Schmidt}, H., {Koester}, D., \& {Allard}, N. 1997, \aap, 325, 1055

\bibitem[{{Venner} {et~al.}(2023){Venner}, {Limbach}, {Vanderburg}, {Blouin}, {Janson}, {Morley}, \& {Stevenson}}]{2023jwst.prop.3621V}
{Venner}, A., {Limbach}, M.~A., {Vanderburg}, A., {et~al.} 2023, {Confirming a Giant Planet Around the White Dwarf GD 140}, JWST Proposal. Cycle 2, ID. \#3621

\bibitem[{{Venner} {et~al.}(2021){Venner}, {Vanderburg}, \& {Pearce}}]{Venner2021}
{Venner}, A., {Vanderburg}, A., \& {Pearce}, L.~A. 2021, \aj, 162, 12, \dodoi{10.3847/1538-3881/abf932}

\bibitem[{Veras(2021)}]{veras2021planetary}
Veras, D. 2021, Planetary Systems Around White Dwarfs.
\newblock \doarXiv{2106.06550}

\bibitem[{{Veras} {et~al.}(2024){Veras}, {Mustill}, \& {Bonsor}}]{2024RvMG...90..141V}
{Veras}, D., {Mustill}, A.~J., \& {Bonsor}, A. 2024, Reviews in Mineralogy and Geochemistry, 90, 141, \dodoi{10.2138/rmg.2024.90.05}

\bibitem[{{Veras} {et~al.}(2016){Veras}, {Mustill}, {G{\"a}nsicke}, {Redfield}, {Georgakarakos}, {Bowler}, \& {Lloyd}}]{2016MNRAS.458.3942V}
{Veras}, D., {Mustill}, A.~J., {G{\"a}nsicke}, B.~T., {et~al.} 2016, \mnras, 458, 3942, \dodoi{10.1093/mnras/stw476}

\bibitem[{{Vigan} {et~al.}(2021){Vigan}, {Fontanive}, {Meyer}, {Biller}, {Bonavita}, {Feldt}, {Desidera}, {Marleau}, {Emsenhuber}, {Galicher}, {Rice}, {Forgan}, {Mordasini}, {Gratton}, {Le Coroller}, {Maire}, {Cantalloube}, {Chauvin}, {Cheetham}, {Hagelberg}, {Lagrange}, {Langlois}, {Bonnefoy}, {Beuzit}, {Boccaletti}, {D'Orazi}, {Delorme}, {Dominik}, {Henning}, {Janson}, {Lagadec}, {Lazzoni}, {Ligi}, {Menard}, {Mesa}, {Messina}, {Moutou}, {M{\"u}ller}, {Perrot}, {Samland}, {Schmid}, {Schmidt}, {Sissa}, {Turatto}, {Udry}, {Zurlo}, {Abe}, {Antichi}, {Asensio-Torres}, {Baruffolo}, {Baudoz}, {Baudrand}, {Bazzon}, {Blanchard}, {Bohn}, {Brown Sevilla}, {Carbillet}, {Carle}, {Cascone}, {Charton}, {Claudi}, {Costille}, {De Caprio}, {Delboulb{\'e}}, {Dohlen}, {Engler}, {Fantinel}, {Feautrier}, {Fusco}, {Gigan}, {Girard}, {Giro}, {Gisler}, {Gluck}, {Gry}, {Hubin}, {Hugot}, {Jaquet}, {Kasper}, {Le Mignant}, {Llored}, {Madec}, {Magnard}, {Martinez}, {Maurel}, {M{\"o}ller-Nilsson}, {Mouillet}, {Moulin}, {Orign{\'e}},
  {Pavlov}, {Perret}, {Petit}, {Pragt}, {Puget}, {Rabou}, {Ramos}, {Rickman}, {Rigal}, {Rochat}, {Roelfsema}, {Rousset}, {Roux}, {Salasnich}, {Sauvage}, {Sevin}, {Soenke}, {Stadler}, {Suarez}, {Wahhaj}, {Weber}, \& {Wildi}}]{Vigan2021}
{Vigan}, A., {Fontanive}, C., {Meyer}, M., {et~al.} 2021, \aap, 651, A72, \dodoi{10.1051/0004-6361/202038107}

\bibitem[{{Wenger} {et~al.}(2000){Wenger}, {Ochsenbein}, {Egret}, {Dubois}, {Bonnarel}, {Borde}, {Genova}, {Jasniewicz}, {Lalo{\"e}}, {Lesteven}, \& {Monier}}]{2000A&AS..143....9W}
{Wenger}, M., {Ochsenbein}, F., {Egret}, D., {et~al.} 2000, \aaps, 143, 9, \dodoi{10.1051/aas:2000332}

\bibitem[{{Werner} {et~al.}(2004){Werner}, {Roellig}, {Low}, {Rieke}, {Rieke}, {Hoffmann}, {Young}, {Houck}, {Brandl}, {Fazio}, {Hora}, {Gehrz}, {Helou}, {Soifer}, {Stauffer}, {Keene}, {Eisenhardt}, {Gallagher}, {Gautier}, {Irace}, {Lawrence}, {Simmons}, {Van Cleve}, {Jura}, {Wright}, \& {Cruikshank}}]{2004ApJS..154....1W}
{Werner}, M.~W., {Roellig}, T.~L., {Low}, F.~J., {et~al.} 2004, ApJs, 154, 1, \dodoi{10.1086/422992}

\bibitem[{{Xu} {et~al.}(2019){Xu}, {Dufour}, {Klein}, {Melis}, {Monson}, {Zuckerman}, {Young}, \& {Jura}}]{2019AJ....158..242X}
{Xu}, S., {Dufour}, P., {Klein}, B., {et~al.} 2019, \aj, 158, 242, \dodoi{10.3847/1538-3881/ab4cee}

\bibitem[{{Xu} {et~al.}(2015){Xu}, {Jura}, {Pantoja}, {Klein}, {Zuckerman}, {Su}, \& {Meng}}]{2015ApJ...806L...5X}
{Xu}, S., {Jura}, M., {Pantoja}, B., {et~al.} 2015, ApJL, 806, L5, \dodoi{10.1088/2041-8205/806/1/L5}

\bibitem[{{Yamaguchi} {et~al.}(2024){Yamaguchi}, {El-Badry}, {Fuller}, {Latham}, {Cargile}, {Mazeh}, {Shahaf}, {Bieryla}, {Buchhave}, \& {Hobson}}]{2024MNRAS.52711719Y}
{Yamaguchi}, N., {El-Badry}, K., {Fuller}, J., {et~al.} 2024, \mnras, 527, 11719, \dodoi{10.1093/mnras/stad4005}

\bibitem[{{Yang} \& {Ishiguro}(2015)}]{2015ApJ...813...87Y}
{Yang}, H., \& {Ishiguro}, M. 2015, \apj, 813, 87, \dodoi{10.1088/0004-637X/813/2/87}

\bibitem[{{Zhan} {et~al.}(2024){Zhan}, {Koll}, \& {Ding}}]{2024arXiv240603189Z}
{Zhan}, R., {Koll}, D. D.~B., \& {Ding}, F. 2024, arXiv e-prints, arXiv:2406.03189, \dodoi{10.48550/arXiv.2406.03189}

\bibitem[{{Zhu} \& {Dong}(2021)}]{2021ARA&A..59..291Z}
{Zhu}, W., \& {Dong}, S. 2021, ARAA, 59, 291, \dodoi{10.1146/annurev-astro-112420-020055}

\bibitem[{{Zuckerman} \& {Becklin}(1987)}]{1987Natur.330..138Z}
{Zuckerman}, B., \& {Becklin}, E.~E. 1987, Nature, 330, 138, \dodoi{10.1038/330138a0}

\bibitem[{{Zuckerman} {et~al.}(2010){Zuckerman}, {Melis}, {Klein}, {Koester}, \& {Jura}}]{2010ApJ...722..725Z}
{Zuckerman}, B., {Melis}, C., {Klein}, B., {Koester}, D., \& {Jura}, M. 2010, ApJ, 722, 725, \dodoi{10.1088/0004-637X/722/1/725}

\end{thebibliography}
\bibliographystyle{aasjournal}
\end{document}